\PassOptionsToPackage{table}{xcolor}
\documentclass[sigconf, nonacm]{acmart}

\setcopyright{none}
\acmConference[Conference acronym 'XX]{Make sure to enter the correct
  conference title from your rights confirmation email}{June 03--05,
  2018}{Woodstock, NY}
\acmDOI{XXXXXXX.XXXXXXX}
\acmISBN{978-1-4503-XXXX-X/2018/06}

\usepackage{CJKutf8}
\usepackage{tabularx}
\usepackage{nicefrac}
\usepackage{makecell}
\usepackage{tcolorbox}
\usepackage{pifont}
\usepackage{alltt}
\usepackage{enumitem}
\usepackage{wrapfig}
\usepackage{float}
\usepackage{algorithm}
\usepackage{algorithmic}
\usepackage{multirow}
\usepackage{nicematrix}
\usepackage{marvosym} 

\newcommand{\ie}{\emph{i.e., }}
\newcommand{\eg}{\emph{e.g., }}
\newcommand{\cf}{\emph{cf. }}

\tcbuselibrary{skins,breakable}
\definecolor{msftBlack}{RGB}{0,0,0}
\newtcolorbox{findingBox}{
    enhanced,
    colback=msftBlack!05,
    colframe=msftBlack!10,
    arc=1mm,
    boxrule=0.5pt,
    left=2mm, right=2mm, top=2mm, bottom=2mm,
    drop fuzzy shadow,
    fontupper=\em,
    notitle
}

\definecolor{panelBack}{HTML}{F1F4F7}
\definecolor{panelRule}{RGB}{130,136,145}
\newtcolorbox{topmatterBox}{
    enhanced,
    colback=panelBack,
    colframe=panelBack,
    boxrule=0pt,
    frame hidden,
    arc=3mm,
    boxsep=0pt,
    left=6mm, right=6mm, top=5mm, bottom=4mm,
    before skip=4mm, after skip=0mm,
    notitle
}
\newcommand{\headerlogo}{%
  \includegraphics[trim={32.8 43.5 23.4 42.3}, clip,
                   height=4mm]{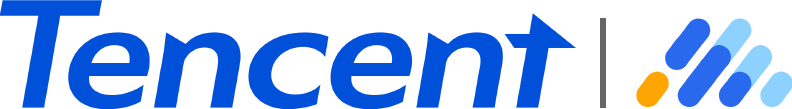}}
\newcommand{\panelnamefont}{\normalsize\bfseries}
\newcommand{\panelaffilfont}{\small}
\newcommand{\panelmailfont}{\ttfamily\footnotesize}
\makeatletter
\newbox\DASH@titlebx
\let\@mktitle\@mktitle@i
\def\@titlefont{\Huge\bfseries}
\def\@mkauthors{\global\setbox\DASH@titlebx=\box\mktitle@bx}
\def\@mkteasers{%
  \global\setbox\mktitle@bx=\vbox{%
    \hsize=\textwidth \linewidth=\textwidth \columnwidth=\textwidth
    \noindent\headerlogo\par
    \vskip 1.25mm
    {\color{panelRule}\hrule height 1.2pt}%
    \vskip 4.75mm
    \box\DASH@titlebx
    \begin{topmatterBox}
      \setlength{\parindent}{0pt}%
      \authorpanel
      \medskip
      \noindent\ignorespaces\@abstract\par
      \ifx\@keywords\@empty\else
        \medskip\noindent{\bfseries\keywordsname:} \@keywords\par
      \fi
      \medskip
      {\color{panelRule}\hrule height 0.4pt}%
      \smallskip
      \noindent{\small\authornotepanel}\par
    \end{topmatterBox}
    \vskip 3.5mm 
  }%
  \global\let\DASH@concepts\@concepts
  \global\let\DASH@keywords\@keywords
  \global\let\@mkabstract\@empty
  \global\@ACM@printccsfalse
  \global\let\@concepts\@empty
  \global\let\@keywords\@empty}
\let\DASH@printendtopmatter\@printendtopmatter
\def\@printendtopmatter{%
  \global\let\@concepts\DASH@concepts
  \global\let\@keywords\DASH@keywords
  \hypersetup{pdfsubject={\@concepts}, pdfkeywords={\@keywords}}%
  \DASH@printendtopmatter}
\makeatother

\title[DASH]{Beyond Action Imitation: Learning a Decision-Aware User Simulator for Online Advertising
}

\newcommand{\authorpanel}{%
  {\panelnamefont
    \mbox{Zipeng Chen$^{1,*}$,}
    \mbox{Jiaer Zheng$^{1,*}$,}
    \mbox{Xiangyang Xu$^{1,*}$,}
    \mbox{Xinyu Lin$^{2}$\textsuperscript{\Letter},}
    \mbox{Zhaobin Wang$^{1}$\textsuperscript{\Letter},}
    \mbox{Zhaohui Liu$^1$,}
    \mbox{Qianjin Xiang$^1$,}
    \mbox{Xiaoyu Zhao$^1$,}
    \mbox{Zhuozhen Yu$^1$,}
    \mbox{Guangshuo Wang$^1$,}
    \mbox{Daxing Chen$^1$,}
    \mbox{Junwei Pan$^1$,}
    \mbox{Zhangbin Zhu$^1$,}
    \mbox{Chengguo Yin$^1$,}
    \mbox{Hao Chen$^1$,}
    \mbox{Tat-Seng Chua$^2$,}
    \mbox{Haijie Gu$^1$,}
    \mbox{Jie Jiang$^1$}\par}
  \smallskip
  {\panelaffilfont
    \mbox{$^1$Tencent Inc., China}\quad
    \mbox{$^2$National University of Singapore, Singapore}\par}
  {\panelmailfont
    $^1$\{beckzpchen, jedizheng, xiangyangxu, zhaobinwang, danielzhliu, randyxiang, xiaoyuzhao, jothanyu,\\
    \hphantom{$^1$}leroygswang, daxingchen, jonaspan, defyzhu, turingyin, haochenchen, jerrickgu, zeus\}@tencent.com\\
    $^2$xylin1028@gmail.com, dcscts@nus.edu.sg\par}%
}
\newcommand{\authornotepanel}{%
  $^*$\,Equal contribution. \quad
  \textsuperscript{\Letter}\,Corresponding author.}

\renewcommand{\shortauthors}{Chen et al.}

\makeatletter
\gdef\authors{Zipeng Chen, Jiaer Zheng, Xiangyang Xu, Xinyu Lin, Zhaobin Wang, Zhaohui Liu, Qianjin Xiang, Xiaoyu Zhao, Zhuozhen Yu, Guangshuo Wang, Daxing Chen, Junwei Pan, Zhangbin Zhu, Chengguo Yin, Hao Chen, Tat-Seng Chua, Haijie Gu, Jie Jiang}
\makeatother

\begin{abstract}
Recent advances in LLM-based user simulation have shown promise for offline evaluation of recommendation and advertising systems. However, existing simulators typically infer user preferences from single-domain interaction histories and are primarily optimized to reproduce observable actions such as clicks. 
Consequently, they capture only a partial view of user preferences, while action-only prediction easily induces model shortcuts and limits both the fidelity and diagnostic value of simulation. 
To address these challenges, we propose \textbf{DASH}, a decision-aware user simulator that jointly \emph{generates thinking traces} and predicts behavioral actions from \emph{heterogeneous cross-domain histories}. 
DASH first introduces a Context Engineering stage that folds heterogeneous cross-domain histories into decision-relevant context, together with prompt optimization for effective reasoning over the folded context. To train a user simulator, DASH distills thinking trajectories from strong LLMs as SFT data, and further tailors a rubric-based reward model that evaluates thinking traces along form, content, and logic for RL training. Combined with the action reward, these signals jointly improve action prediction and thinking quality. Extensive experiments on real-world Tencent advertising data spanning five heterogeneous content domains demonstrate the effectiveness, efficiency, fidelity, and diagnostic value of DASH.

\end{abstract}



\begin{document}

\maketitle

\section{Introduction}
\label{sec:intro}

Online advertising systems have achieved remarkable commercial success, with advances in deep learning and multi-stage ranking substantially improving click-through rate prediction and conversion optimization~\cite{goldfarb2011online, cheng2016wide, covington2016deep, zhou2018deep, guo2017deepfm}. Yet advertising strategies are becoming increasingly complex (\eg expanded feature space of ads), which makes live A/B testing for evaluating candidate algorithms prohibitively costly and may also degrade user experience~\cite{kohavi2020trustworthy, shi2019virtual}.
To enable low-risk experimentation and scalable offline evaluation, building user simulators to serve as reliable proxies of real users has become a promising avenue and has attracted growing attention from both academia and industry~\cite{shi2019virtual,zhang2024agent4rec,ABAgent,liu2025recoworld,lin2026autonomous}. 

User simulation typically involves two key steps: modeling user preferences from historical interactions, and then predicting the user’s response to a target item. Early approaches rely primarily on rule-based methods~\cite{shi2019virtual,zhu2024reliable,wang2023rethinking} or Reinforcement Learning (RL) environments~\cite{zhao2023kuaisim,ie2019recsim}. 
More recent work uses LLM-based agents to simulate users by introducing profile, memory, and reflection mechanisms~\cite{wang2025user,zhang2024agent4rec,recusersim,ABAgent}, or fine-tunes LLMs with logged user behaviors~\cite{wei2025mirroring,chen2026vragent,zhang2025shop,wang2025customer}. Despite this progress, existing user simulators suffer from limitations in both steps:
\begin{itemize}[leftmargin=1em, itemsep=0pt, topsep=0pt, parsep=0pt, partopsep=0pt]

    \item \textbf{Single-domain preference modeling.}
    Existing approaches infer user preferences from interaction histories within a single domain. However, in the real world, users interact with content across multiple domains, which jointly shape their responses to a target item. 
    Consequently, single-domain histories provide only a partial view of user preferences, limiting the simulator's ability to accurately predict user responses.

    \item \textbf{Action-only response modeling.} 
    Most existing methods merely predict the user's final actions (\eg clicks). 
    However, action-only prediction easily induces model shortcuts (\eg overfitting to specific actions), yielding biased and inaccurate responses~\cite{chen2026towards,zhu2026realusersim}.
    Moreover, while predicted actions can directly evaluate recommender models, they offer little signal on why the models fail to satisfy users.  
    As such, action-only prediction limits user simulators in both simulation fidelity and diagnostic value.

\end{itemize}
Based on the above insights, a good user simulator should capture heterogeneous real-world interactions while reliably modeling the underlying decision traces for diagnostic insights. 
Nonetheless, achieving this is non-trivial with two key challenges: 

\begin{itemize}[leftmargin=1em, itemsep=0pt, topsep=0pt, parsep=0pt, partopsep=0pt]
    \item \textbf{\emph{Effective Heterogeneous Context Folding} (\(\mathbf{C}_1\)):} it is essential to incorporate cross-domain history for accurate simulation, because user preference is collectively reflected in heterogeneous behaviors. 
    However, such histories might contain irrelevant interactions that mislead the simulation, and can easily exceed the context budget of LLMs. 
    Therefore, it is crucial to fold heterogeneous context into decision-relevant ones. 

    \item \textbf{\emph{Effective Supervision for Thinking Traces} (\(\mathbf{C}_2\)):} 
    fine-tuning on user behavior data is necessary for LLMs to adapt from pre-training tasks to user simulation~\cite{wei2025mirroring,chen2026vragent,wang2025customer}. 
    However, unlike observable actions (\eg clicks), users' underlying thinking traces are not naturally available in recommendation data, leaving no ground truth for direct \emph{Supervised Fine-Tuning}~(SFT). 
    While RL could bypass thinking traces' exact token supervision, it instead requires reward signals to assess thinking quality, but such reward signals are also unavailable due to the lack of evaluation approach. 
    It is therefore crucial to construct effective supervision signals for thinking traces. 
\end{itemize}

In light of this, we propose \textbf{DASH}, a \textbf{D}ecision-\textbf{A}ware \textbf{S}imulator with \textbf{H}eterogeneous context, for modeling both user thinking traces and behavioral actions in advertising. \textbf{To address \(\mathbf{C}_1\)}, 
we introduce \emph{Context Engineering}~(CE) stage, which aims to fold the noisy context into an informative one. Specifically, we tailor a compression strategy for different interaction types (\eg preserve all descriptions for ``click'' and retain only category information for ``skip''). To balance across different domains (\eg ads, video content), we allocate domain-specific context budgets based on informativeness. We further design a prompt optimization strategy to facilitate effective reasoning over folded context. 
\textbf{To address \(\mathbf{C}_2\)}, we construct both token-level and trace-level supervision signals for training a small model\footnote{Practical advertising systems require an efficient model to satisfy latency constraints for online serving, whereas frontier LLMs incur prohibitive inference costs.} via the \emph{SFT-then-RL paradigm}~\cite{zhang2025shop,wang2025customer,wei2025mirroring}. 
Specifically, we use strong LLMs to generate thinking trajectories as SFT data (token-level), and further tailor a rubric-based reward model that evaluates thinking traces along various dimensions (\ie form, content, and logic at the trace level) for RL. 
Combined with the action reward, these signals improve both action prediction and thinking quality. 
Our key contributions are summarized as follows:

\begin{itemize}[leftmargin=1em, itemsep=0pt, topsep=0pt, parsep=0pt, partopsep=0pt]
    \item We identify two critical limitations of existing user simulators, \ie single-domain preference modeling provides only a partial view of user preferences, while action-only response modeling undermines both the action fidelity and diagnostic value of user simulation in practice.
    \item We propose DASH, a decision-aware user simulator that captures heterogeneous interactions via the CE stage and models users' thinking traces through the SFT and RL stages with tailored token-level and trace-level supervision, enabling both accurate action prediction and diagnostic simulation.
    \item We conduct experiments on real-world Tencent advertising data combining advertising interactions with native behaviors from five heterogeneous content domains. Empirical results demonstrate the effectiveness, efficiency, fidelity, and diagnostic value of the proposed user simulator. 
\end{itemize}

\section{Related Work}
\label{sec:related_work}

\subsection{User Simulation for Recommendation}
Traditional user simulators rely mainly on rule-based methods, generative models, or reinforcement learning environments. Representative systems, including RecSim, Virtual-Taobao, RecoGym, and KuaiSim~\cite{ie2019recsim,shi2019virtual,rohde2018recogym,zhao2023kuaisim}, support controllable policy evaluation by modeling observable actions such as clicks, skips, and purchases, but provide limited insight into users' underlying decision processes.
Recent LLM-based simulators, such as Agent4Rec, RecAgent, RecUserSim, and RecoWorld introduce profile, memory, and reflection mechanisms to simulate more personalized user behaviors~\cite{wang2025user,recusersim,liu2025recoworld,zhang2024agent4rec,CSHI,ABAgent}. Subsequent studies further align simulators with real user behavior through fine-tuning~\cite{wei2025mirroring,chen2026vragent,xu2026unveiling,zhang2025shop,wang2025customer}. Despite these advances, existing LLM-based simulators are primarily optimized for action-level behavior imitation. They often rely on single-domain user histories and action-only supervision, limiting both user preference modeling and thinking traces modeling. In contrast, our work models both user actions and their underlying thinking traces from heterogeneous contexts in advertising.

\subsection{LLM Reasoning and Alignment}

Recent advances in LLMs have improved multi-step reasoning and controllable generation. Chain-of-thought prompting and self-refinement methods enable models to generate and iteratively improve intermediate reasoning traces~\cite{wei2022chain,yao2023tree,madaan2023selfrefine,shinn2023reflexion}. Meanwhile, post-training techniques such as RLHF, RLAIF, and rubric-based reinforcement learning align model behavior using human preferences, AI feedback, and structured rubrics~\cite{ouyang2022training,bai2022constitutional,gunjal2025rubrics}. Existing LLM-based reasoning and alignment methods in recommendation have primarily focused on improving preference understanding and recommendation generation~\cite{fang2025reason4rec,lin2026bringing,lin2026verifiable,you2026r}. However, these methods typically use reasoning for improving recommendation performance, rather than explicitly modeling and supervising user thinking traces. In contrast, our work adapts reasoning and alignment techniques to advertising user simulation, jointly optimizing thinking traces and behavioral actions.
\section{Problem Formulation}

\label{sec:simulation}
We formulate simulation tasks spanning thinking and action prediction as a unified next-token prediction problem. Formally, we construct the input $X = (\mathcal{I},\, c_t,\, d_t,\, \mathcal{C})$ with four components: the task instruction $\mathcal{I}$, the ad request context $c_t$, the ad profile $d_t$, and the user context $\mathcal{C} = (p_u,\, \mathcal{H}_u^{\text{ad}},\, \mathcal{H}_u^{\text{content}})$ comprising the user profile $p_u$, the ad interaction history $\mathcal{H}_u^{\text{ad}}$, and the heterogeneous content behavior history $\mathcal{H}_u^{\text{content}}$. The action space is $\mathcal{A} =\{\textit{skip},\allowbreak\, \textit{click},\allowbreak\, \textit{conversion},\allowbreak\, \textit{negative-feedback}\}$. Given $X$, the model maximizes the likelihood of the target response $Y$:
\begin{equation}\small
  \mathcal{L}(\theta) = -\sum_{i=1}^{|Y|} \log P_\theta(y_i \mid X,\, y_{<i}),
\end{equation}
where $Y = (\tau,\; a)$ is the simulated user response: $\tau$ denotes the thinking trace, a sequence that explicates the user's decision process; and $a \in \mathcal{A}$ denotes the predicted behavioral action. 
\section{DASH}
\label{sec:method}

\begin{figure*}[t]
\centering
\captionsetup{belowskip=0pt, aboveskip=0pt}
\includegraphics[width=\textwidth]{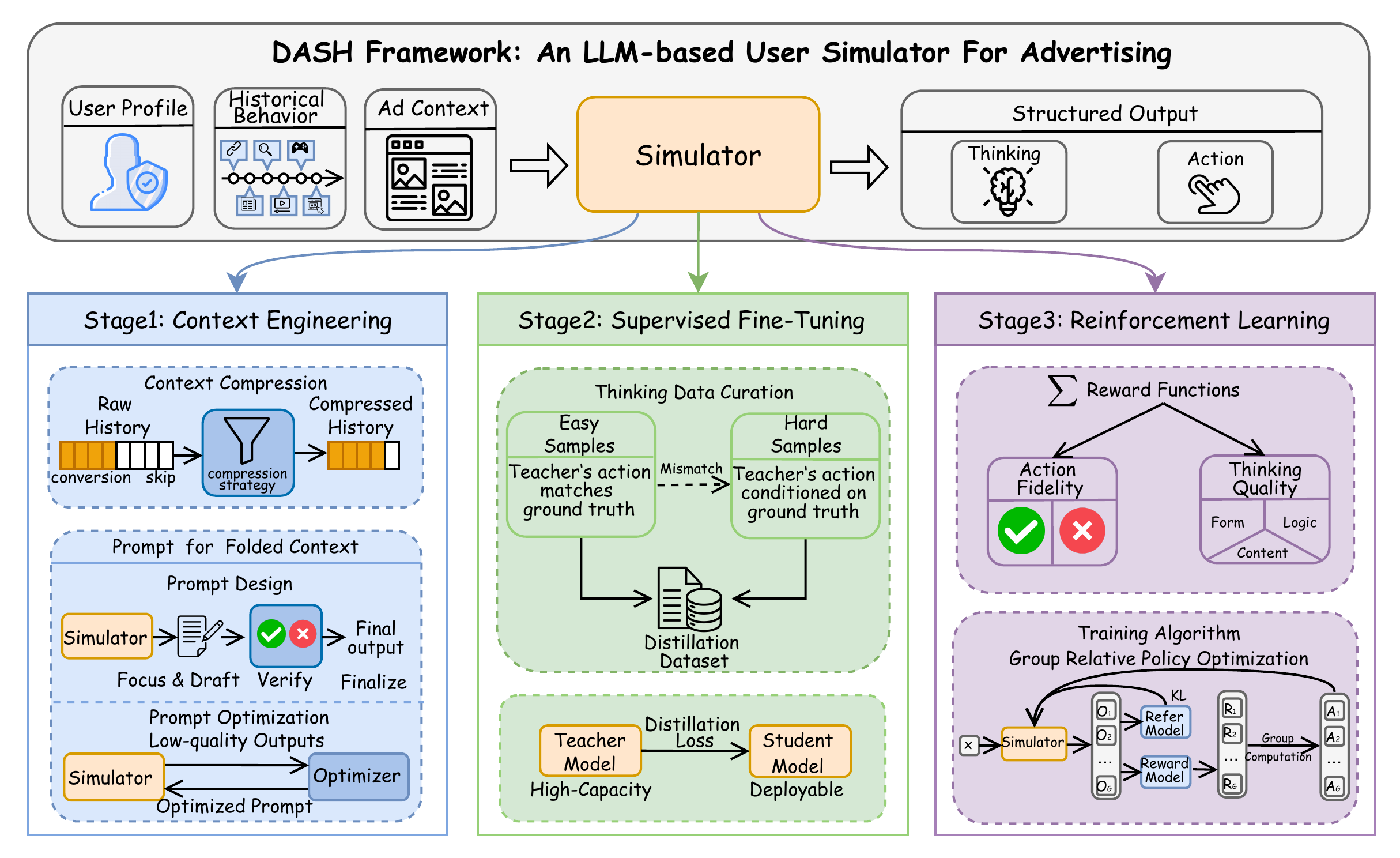}
\caption{Overview of DASH, a progressive CE-SFT-RL framework for advertising user simulation. Stage~1 uses context compression and prompt optimization to leverage decision-relevant signals for effective reasoning. Stage~2 constructs token-level supervision signals, distilling quality-filtered teacher trajectories into a student model. Stage~3 further improves action fidelity and thinking quality through GRPO with a hybrid trace-level reward (\ie action fidelity and thinking quality).}
\label{fig:framework}
\end{figure*}

This work proposes {DASH}, which jointly generates user thinking traces and predicts behavioral actions, as shown in Figure~\ref{fig:framework}. 
We set three consecutive objectives to pursue the thinking simulation: 
1) \textit{heterogeneous context folding}, which leverages informative context that could reflect underlying thinking process; 
2) \textit{token-level optimization}, which performs SFT on teacher trajectories to acquire basic thinking generation capability; and
3) \textit{trace-level optimization}, which applies RL with trace-level rewards to improve simulation performance by addressing the limited coverage of teacher.
Accordingly, we build DASH with a progressive three-stage framework following a \emph{Context Engineering $\rightarrow$ Supervised Fine-Tuning $\rightarrow$ Reinforcement Learning} paradigm, with each stage building on the capabilities acquired in the preceding stage.  

\begin{figure}[t]
\centering
\captionsetup{belowskip=0pt, aboveskip=0pt}
\includegraphics[width=\columnwidth]{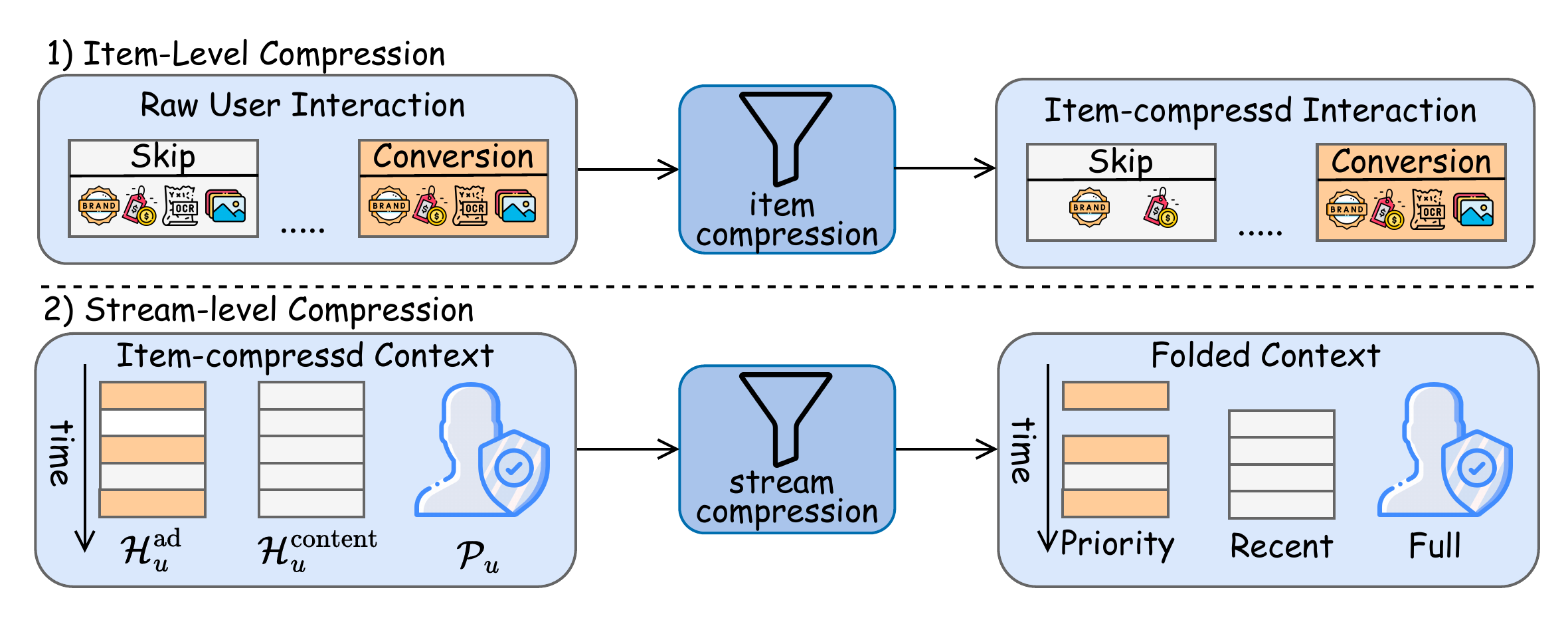}
\caption{Hierarchical context compression for preserving decision-relevant signals from heterogeneous user histories through item-level and stream-level compression.}
\label{fig:compression}
\end{figure}

\subsection{Stage~1: Context Engineering}
\label{sec:stage1}

This stage aims to achieve more effective context folding via user history compression and task prompt optimization. 
1) To fold heterogeneous user histories into a compact context, we introduce a hierarchical context compression strategy, which aims to preserve sparse, decision-relevant signals under a limited context budget. 
2) To provide an effective prompt, we design a structured self-refinement prompting strategy and optimize it through a closed-loop prompt optimization process, which aims to further facilitate reasoning over folded context. 

\subsubsection{\textbf{Hierarchical Context Compression.}}
\label{sec:compression}

In real-world applications, heterogeneous user histories can be extremely long, reaching millions of tokens in industrial settings, while decision-relevant signals are often sparse and buried in noisy behaviors. Meanwhile, applying additional retrieval or generative summarization may introduce considerable computational and latency overhead. To compress such long and noisy histories both efficiently and effectively, we introduce a hierarchical compression strategy that operates at item level and stream level, as illustrated in Figure~\ref{fig:compression}.

\textbf{\textit{Item-Level Compression.}}
Different action types convey different amounts of preference information: frequent skips provide relatively weak signals, whereas less frequent clicks, conversions, and negative-feedback are more informative. We therefore tailor the compression strategy to each action type. For low-information interactions with $a_i \in \{\textit{skip}\}$, we retain only category-level metadata of the item profile. For high-information interactions with $a_i \in \{\textit{click},\allowbreak\, \textit{conversion},\allowbreak\, \textit{negative-feedback}\}$, we preserve the full item profile. This compression effectively prevents sparse but high-information signals of users' latent preferences from being overwhelmed by frequent skips.

\textbf{\textit{Stream-Level Compression.}}
The context consists of three components with distinct token footprints: the user profile $p_u$, the advertising history $\mathcal{H}_u^{\text{ad}}$, and the content history $\mathcal{H}_u^{\text{content}}$. Even after substantial item-level compression, the two behavioral histories still account for most of the input. In real-world interaction logs, they can collectively span millions of tokens and exceed the effective context window of contemporary LLMs (typically $\leq$ 32K tokens)~\cite{hsieh2024ruler,liu2024lost}. We therefore retain the user profile in full, fill the advertising stream by prioritizing high-information interactions over low-information ones, and truncate the content stream to its most recent interactions to capture the user's current preference state. To ensure that the resulting context remains within the context limit, we assign separate token quotas $q_{\text{profile}}$, $q_{\text{ad}}$, and $q_{\text{content}}$ to the three components, subject to
\begin{equation}\small
q_{\text{profile}} + q_{\text{ad}} + q_{\text{content}} \leq L_{\max},
\end{equation}
where $L_{\max}$ denotes the target context length.

\begin{figure}[t]
\vspace{-0.2cm}
\centering
\captionsetup{belowskip=0pt, aboveskip=0pt}
\includegraphics[width=\columnwidth]{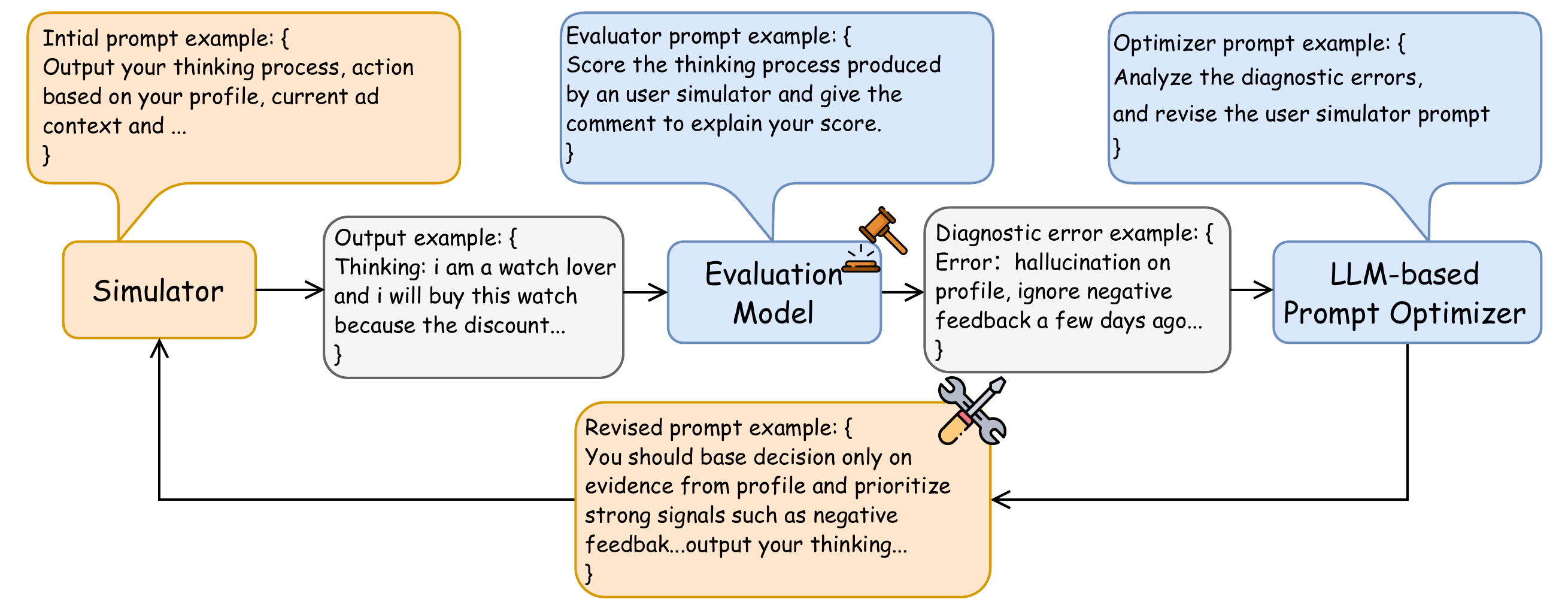}
\caption{Prompt optimization: simulator outputs are evaluated for action correctness and thinking quality, and identified errors are analyzed by an LLM-based optimizer to revise the prompt for the next iteration.}
\label{fig:prompt_optim}
\end{figure}

\subsubsection{\textbf{Prompt for Folded Context.}}
\label{sec:prompt_design}
Although hierarchical context compression preserves decision-relevant signals within the context limit, the simulator may still fail to effectively reason over the folded context~(\eg by generating unsupported statements). We therefore improve the prompt to facilitate effective reasoning over the folded context in two ways: 1) designing a structured self-refinement prompt; and 2) optimizing it through a closed-loop process based on the generated simulation outputs.

\textbf{\textit{Prompt Design.}}
Inspired by recent advances in agentic refinement~\cite{fu2025agentrefine,madaan2023selfrefine,shinn2023reflexion}, we design a structured self-refinement prompt to guide the simulator in reasoning over the folded context, \ie a reasoning process including \textit{focus}, \textit{draft}, \textit{verify}, and \textit{finalize}. 
In \textit{Focus}, the simulator identifies decision-relevant signals from the folded context with respect to the current ad. In \textit{Draft}, it generates an initial thinking trace and a candidate action based on the identified signals. In \textit{Verify}, it checks whether the claims in the thinking trace are supported by the folded context and whether the candidate action follows logically from the trace. In \textit{Finalize}, it revises the draft if unsupported claims or logical inconsistencies are detected; otherwise, it retains the original output.

\textbf{\textit{Prompt Optimization.}}
Although the self-refinement strategy provides an explicit reasoning workflow, a manually designed prompt may not generalize well across the combinatorial diversity of user profiles, behavioral histories, and advertising contexts. 
To address this limitation, drawing on recent advances in prompt engineering~\cite{zhou2022large,yang2023large,pryzant2023automatic}, we design a prompt optimization process, as illustrated in Figure~\ref{fig:prompt_optim}. Starting from an initial self-refinement prompt, the simulator generates a thinking trace and an action for each sample. Outputs with incorrect actions or errors in their thinking traces are retained as error cases. An LLM-based prompt optimizer analyzes recurring failure patterns across these cases and revises the prompt accordingly. The revised prompt is then used in the next iteration, allowing the prompt to progressively improve factual grounding and the logical consistency. 

\subsection{Stage~2: Supervised Fine-Tuning}
\label{sec:stage2}

\begin{figure}[t]
\vspace{-0.2cm}
\centering
\captionsetup{belowskip=0pt, aboveskip=0pt}
\includegraphics[width=\columnwidth]{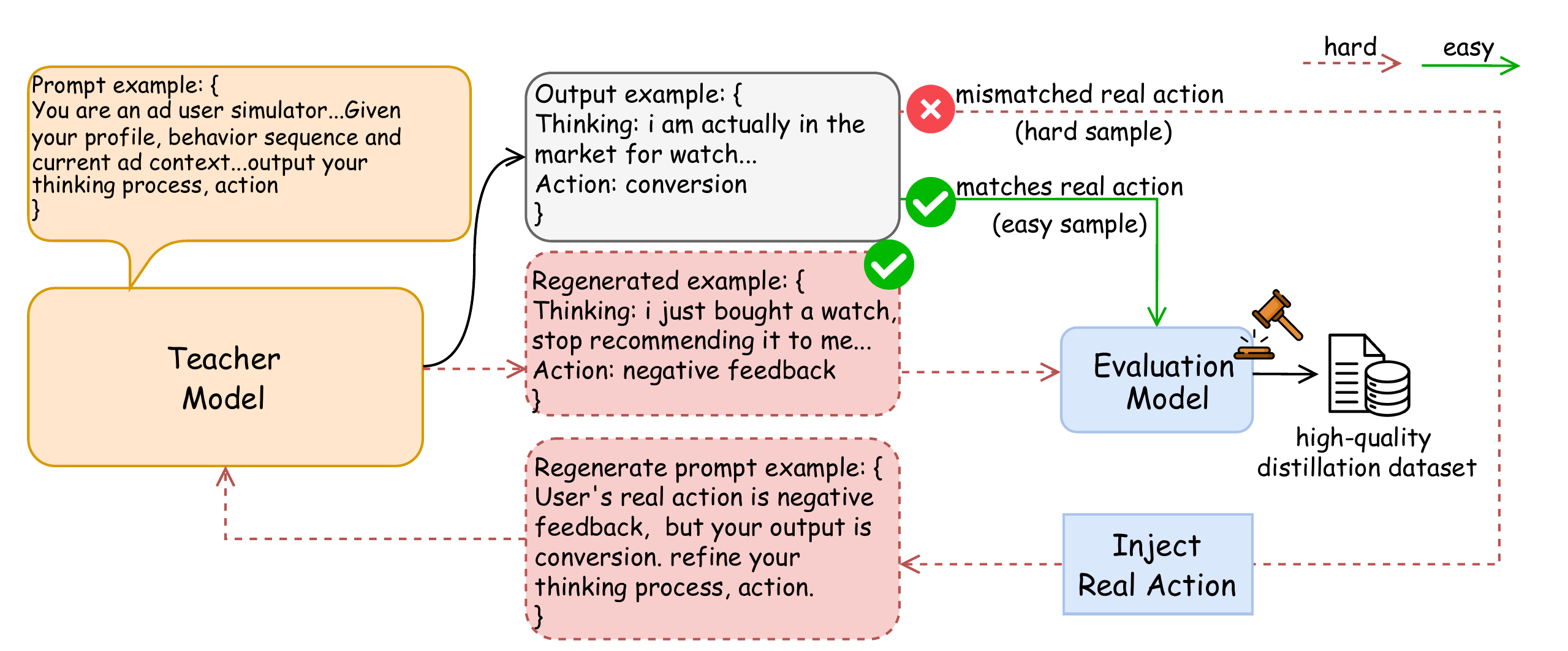}
\caption{SFT data curation pipeline: correct teacher predictions are collected as easy samples; mismatched cases are regenerated into hard samples, and both are quality-filtered into the final training set.}
\label{fig:sft_data_curation}
\end{figure}
With the compressed context, the model still lacks task-specific knowledge, limiting its simulation performance (as shown in Section~\ref{sec:sft_analysis}).
This stage therefore aims to construct users' thinking traces for token-level model optimization, thereby injecting basic thinking generation capability for the model via SFT~\cite{guo2025deepseek}.

\subsubsection{\textbf{Training Data Curation}}
\label{sec:data_curation}

\begin{table*}[t]
\centering
\captionsetup{belowskip=0pt, aboveskip=0pt}
\caption{Hierarchical decomposition of the rubric-based thinking-quality reward
$R_{\text{think}}$. Each dimension is decomposed into sub-dimension rewards and
fine-grained rubric rewards.}
\label{tab:thinking_reward}
\footnotesize
\renewcommand{\arraystretch}{1.2}
\setlength{\tabcolsep}{4pt}

\begin{tabularx}{\textwidth}{@{}l l X l@{}}
\toprule
\textbf{Dimension} &
\textbf{Sub-dimension} &
\textbf{Rubrics} &
\textbf{Scoring} \\
\midrule

\multirow{2}{*}{\makecell[l]{Form\\($R_{\text{form}}$)}}
  & Linguistic Expression ($R_{\text{exp}}$)
  & Grammar Correctness,
    Fluency,
    Naturalness,
    Diversity
  & LLM, 1--5 \\

  & Thinking-step Completeness ($R_{\text{TC}}$)
  & Thinking-step Coverage Score
  & Binary \\

\midrule

\multirow{2}{*}{\makecell[l]{Content\\($R_{\text{content}}$)}}
  & Factual Correctness ($R_{\text{fact}}$)
  & Hallucination Rate,
    Overall Factual Accuracy
  & LLM, 1--5 \\

  & Business-knowledge Alignment ($R_{\text{align}}$)
  & User Profile Alignment,
    Ad Attribute Alignment,
    Scenario Context Alignment
  & LLM, 1--5 \\

\midrule

\multirow{3}{*}{\makecell[l]{Logic\\($R_{\text{logic}}$)}}
  & Attribution Clarity ($R_{\text{attr}}$)
  & Key Factor Coverage,
    Evidence Support
  & LLM, 1--5 \\

  & Thinking Coherence ($R_{\text{coher}}$)
  & Causal Coherence,
    Internal Consistency
  & LLM, 1--5 \\

  & Thinking--Action Alignment ($R_{\text{behav}}$)
  & Behavior Alignment
  & LLM, 1--5 \\

\bottomrule
\end{tabularx}
\end{table*}
We generate both easy and hard training samples to ensure comprehensive coverage, followed by a quality filtering step that retains only high-quality distilled trajectories, as illustrated in Figure~\ref{fig:sft_data_curation}.

\textbf{\textit{Easy Sample Collection.}}
These samples are those in which the teacher model's generated thinking trace naturally aligns with the ground-truth action recorded in real-world logs. Such concordant samples serve as standard imitation learning targets:
\begin{equation}
  \mathcal{D}_{\text{easy}} = \{(x_i,\; \hat{\tau}_i,\; a_i^{*}) \mid \pi_{\text{teacher}}(x_i) \rightarrow \hat{\tau}_i,\; \hat{a}_i = a_i^{*}\},
\end{equation}
where $x_i$ is the input context, $\hat{\tau}_i$ is the teacher-generated thinking trace, and $\hat{a}_i$ and $a_i^*$ are the teacher-generated and ground-truth actions, respectively.

\textbf{\textit{Hard Sample Regeneration.}}
When the teacher's action prediction differs from the ground truth ($\hat{a}_i \neq a_i^{*}$), we construct hard samples by providing the teacher with the ground-truth action and prompting it to regenerate a thinking trace $\tilde{\tau}_i$: 
\begin{equation}
  \mathcal{D}_{\text{hard}} = \{(x_i,\; \tilde{\tau}_i,\; a_i^{*}) \mid \pi_{\text{teacher}}(x_i, a_i^{*}) \rightarrow \tilde{\tau}_i\}.
\end{equation}

\textbf{\textit{Quality Filtering.}} 
The generated thinking traces may contain hallucinated thinking and logically inconsistent content because some thinking traces are generated based on the ground-truth action instead of history (\ie hard samples). 
Therefore, to prevent such low-quality trajectories from propagating to the student, 
we apply a quality filtering step with a judge LLM $\pi_{\text{judge}}$. 
Concretely, for every candidate sample in $\mathcal{D}_{\text{easy}} \cup \mathcal{D}_{\text{hard}}$, the judge LLM evaluates the thinking trace according to the predefined fine-grained rubrics~(detailed in Appendix~\ref{appendix:eval_rubrics}). 
Based on the scores, we only retain the high-score samples exceeding a threshold $\eta_{\text{sft}}$: 
\begin{equation}
  \mathcal{D}_{\text{filtered}} = \left\{(x_i,\; \tau_i,\; a_i^{*}) \in \mathcal{D}_{\text{easy}} \cup \mathcal{D}_{\text{hard}} \;\middle|\; \pi_{\text{judge}}(\tau_i) \geq \eta_{\text{sft}} \right\}.
\end{equation}

\subsubsection{\textbf{Training Objective}}

The student model $\pi_\theta$ is optimized via standard autoregressive language modeling over the quality-filtered dataset $\mathcal{D} = \mathcal{D}_{\text{filtered}}$:
\begin{equation}
  \mathcal{L}_{\text{SFT}} = -\mathbb{E}_{(x, \tau, a) \sim \mathcal{D}} \left[\log \pi_\theta(\tau, a \mid x)\right].
\end{equation}

\subsection{Stage~3: Reinforcement Learning}
\label{sec:stage3}

Although token-level optimization adapts the student model to the simulation task, it relies primarily on imitating each token and therefore remains limited by the coverage of distilled trajectories (empirical evidence is in Section~\ref{sec:experiments}). 
This stage therefore aims to further improve thinking quality and action prediction through trace-level supervision signals via RL. 
However, jointly optimizing these two objectives is non-trivial: action correctness provides only sparse outcome-level feedback, while reward signals for thinking traces remain unavailable due to the lack of evaluation approach. 
To address this challenge, we 
design a hybrid action-thinking reward that augments sparse action feedback with fine-grained assessments of thinking traces. 

\subsubsection{\textbf{Reward Design}}
\label{sec:reward}
Our reward design is inspired by recent rubric-as-reward methods, which use structured rubrics to provide fine-grained supervision for open-ended tasks~\cite{gunjal2025rubrics,liu2025openrubrics}. In our setting, however, simulation quality hinges on two complementary desiderata: alignment with logged action and the quality of the thinking trace. We therefore define a hybrid reward:
\begin{equation}
R_{\text{total}} = R_{\text{action}} + R_{\text{think}},
\label{eq:reward}
\end{equation}
where $R_{\text{action}}$ measures action accuracy against real user logs, and $R_{\text{think}}$ is a rubric-based reward that decomposes quality into form, content, and logic dimensions. We detail each component below.

\textbf{\textit{Action Reward ($R_{\text{action}}$).}} We define the action reward by comparing the predicted action $\hat{a}$ and the ground-truth action $a^{*}$:
\begin{equation}
  R_{\text{action}} = \begin{cases}
    1, & \text{if } \hat{a} = a^{*}, \\
    0, & \text{if } \hat{a} \neq a^{*}.
  \end{cases}
\end{equation}

\textbf{\textit{Thinking Quality Reward ($R_{\text{think}}$).}} 
We introduce a rubric-based thinking reward to provide trace-level supervision. As summarized in Table~\ref{tab:thinking_reward}, we decompose thinking quality into three complementary dimensions: \textit{form}, \textit{content}, and \textit{logic}. Each dimension is represented by a weighted combination of its corresponding sub-dimension rewards:
\begin{align}
  R_{\text{think}} =\; & w_1 \cdot \underbrace{(a_1 R_{\text{exp}} + a_2 R_{\text{TC}})}_{R_{\text{form}}} \;+\; w_2 \cdot \underbrace{(b_1 R_{\text{fact}} + b_2 R_{\text{align}})}_{R_{\text{content}}} \nonumber \\
  & +\; w_3 \cdot \underbrace{(c_1 R_{\text{attr}} + c_2 R_{\text{coher}} + c_3 R_{\text{behav}})}_{R_{\text{logic}}}.
\end{align}
Here, $w_1$, $w_2$, and $w_3$ control the relative importance of the three dimensions, while $a_i$, $b_i$, and $c_i$ denote the weights of their corresponding sub-dimensions. 
Each sub-dimension is aggregated by a group of fine-grained rubrics with different weights. 
For example, the linguistic-expression sub-dimension comprises rubric rewards for grammar correctness, fluency, naturalness, and diversity (detailed descriptions can be found in Appendix~\ref{appendix:eval_rubrics}). 
This hierarchical decomposition provides fine-grained credit assignment for intermediate thinking traces, yielding denser supervision than the sparse outcome-level feedback available from logged actions. It enables RL to distinguish not only whether a final action is correct, but also whether the underlying decision process is well-formed, evidence-grounded, and causally coherent.

To obtain the reward for each dimension,  
we employ an LLM as reward model, following the LLM-as-a-Judge paradigm~\cite{zheng2023judging}. 
The reward model evaluates the generated thinking trace under each predefined rubric and assigns a score on a 1--5 Likert scale, which is normalized to $[0,1]$ before aggregation. 
The complete rubric definitions and weight configurations are provided in Appendix~\ref{appendix:eval_rubrics} and Appendix~\ref{appendix:reward_weights}, respectively.

\subsubsection{\textbf{GRPO Training}}
\label{sec:grpo}

We adopt Group Relative Policy Optimization (GRPO)~\cite{shao2024deepseekmath} as our RL framework. For each input $x$, GRPO samples a group of $G$ candidate outputs $\{o_1, \ldots, o_G\}$ from the current policy $\pi_\theta$. The policy gradient objective is formulated as:
\begin{align}
  \mathcal{L}_{\text{GRPO}}(\theta) = -\frac{1}{G}\sum_{i=1}^{G} \Big[ &\min\!\big( r_i(\theta)\, \hat{A}_i, \text{clip}\!\big(r_i(\theta), 1-\epsilon, 1+\epsilon\big) \hat{A}_i \big) \nonumber \\
  &- \beta \cdot \mathbb{D}_{\mathrm{KL}}(\pi_\theta \| \pi_{\text{ref}}) \Big],
\end{align}
where $r_i(\theta) = \frac{\pi_\theta(o_i \mid x)}{\pi_{\text{old}}(o_i \mid x)}$ is the importance sampling ratio with respect to the policy before the current update step, $\epsilon$ is the clipping ratio, $\pi_{\text{ref}}$ is the frozen reference policy, and $\beta$ controls the KL regularization strength. Following the standard GRPO formulation~\cite{shao2024deepseekmath}, we compute group-relative advantages $\hat{A}_i$ through within-group reward normalization and apply per-token KL regularization against the reference policy $D_{\mathrm{KL}}$.

\section{Experiments}
\label{sec:experiments}
In this section, we conduct experiments to answer the following research questions (RQ):
\begin{itemize}[leftmargin=1em, itemsep=0pt, topsep=0pt, parsep=0pt, partopsep=0pt]
    \item \textbf{RQ1:} How does DASH compare with baselines across different LLMs in terms of action prediction and thinking quality?
    \item \textbf{RQ2:} What are the contributions of the key components (\ie CE, SFT, and RL) to the simulation performance?
    \item \textbf{RQ3:} What are the effects of different context sources and prompt optimization in CE on simulation performance?
    \item \textbf{RQ4:} What are the effects of SFT strategies and RL reward design on simulation performance?
    \item \textbf{RQ5:} How well does the LLM-based evaluator align with human judgments in assessing thinking quality?
\end{itemize}

\subsection{Setup}

\subsubsection{\textbf{Dataset and Metrics.}}
\label{sec:setup}
We conduct experiments on the cross-domain advertising dataset collected from the Tencent platform (Appendix~\ref{appendix:dataset_details}), which contains heterogeneous interaction histories across domains for each user. To prevent information leakage, we strictly split the data by temporal order into training, validation, and test sets, ensuring that all test instances occur chronologically after the training period. All results reported in this section are evaluated on a held-out test set of 1{,}000 samples. 

For evaluation metrics, we evaluate all models from two complementary perspectives: action prediction and thinking quality. For action prediction, we report class-support-weighted Precision, Recall, and F1 score, denoted as W-Prec., W-Rec., and W-F1, respectively. For thinking quality, we report Form, Content, and Logic scores, together with their weighted aggregate, denoted as Think. These dimensions are consistent with the rubric-based reward design described in Section~\ref{sec:reward}. For reporting, all LLM-generated thinking-quality scores are rescaled from $[0,1]$ to $[0,100]$.

\subsubsection{\textbf{Baselines.}}
As we reformulate the user simulation task as a joint modeling of thinking and action prediction, most existing user simulators (\eg action-only prediction) fail to be directly compared under this setting. 
As such, we adopt the representative prompting methods~\cite{wang2023rethinking,zhu2024reliable,kim2024stop} to achieve joint generation of both thinking and action prediction. 
We instantiate the pre-defined prompt on a set of representative open-source LLMs, including DeepSeek-V4-Flash~\cite{deepseekai2026deepseekv4}, Kimi-K2.5~\cite{kimiteam2026kimik25visualagentic}, MiniMax-M2.5~\cite{minimax2025m25}, Qwen3.5-397B-A17B (hereafter Qwen3.5-397B) and Qwen3.5-35B~\cite{qwen3.5}. 
Refer to Appendix~\ref{appendix:prompt_ours} for detailed prompts to achieve joint simulation.

\begin{table}[t]
\centering
\captionsetup{belowskip=0pt, aboveskip=0pt}
\caption{Main results on the DASH evaluation suite. The best/second best scores in each column are \textbf{bolded}/\underline{underlined}. (``Think.'' represents the overall thinking quality score aggregated from three thinking dimensions.)}
\label{tab:result_all}
\setlength{\tabcolsep}{3pt}
\setlength{\extrarowheight}{0pt}
\renewcommand{\arraystretch}{0.95}
\setlength{\aboverulesep}{0pt}
\setlength{\belowrulesep}{0pt}
\resizebox{\columnwidth}{!}{
\begin{tabular}{l ccc cccc}
\toprule
\multirow{2}{*}{\textbf{Model}}
 & \multicolumn{3}{c}{\textbf{Action Quality (\%)}}
 & \multicolumn{4}{c}{\textbf{Thinking Quality (\%)}} \\
\cmidrule(lr){2-4}\cmidrule(lr){5-8}
 & W-Prec. & W-Rec. & W-F1
 & Form & Cont. & Logic & Think. \\
\midrule
Qwen3.5-35B        & 54.31 & 56.50 & 55.18 & 90.20 & 88.86 & 88.99 & 89.07 \\
MiniMax-M2.5~(230B)       & 53.24 & 49.95 & 48.48 & 87.52 & 92.59 & 87.02 & 88.74 \\
DeepSeek-V4-Flash~(284B)  & 56.74 & 56.70 & 55.90 & \underline{93.45} & \textbf{93.44} & 90.45 & 91.65 \\
Qwen3.5-397B       & 59.57 & 61.30 & 60.23 & 92.96 & \underline{93.28} & 90.25 & 91.43 \\
Kimi-K2.5~(1T)          & 57.65 & 57.80 & 56.47 & 92.83 & 91.55 & \underline{91.94} & \underline{91.91} \\
\midrule
\rowcolor{black!7}\multicolumn{8}{l}{\textbf{DASH}~(Qwen3.5-35B)} \\
Small-scale SFT    & \underline{59.69} & 61.30 & 59.07 & 91.61 & 88.52 & 90.64 & 90.10 \\
Large-scale SFT    & 59.49 & \underline{61.70} & \underline{60.48} & 92.07 & 87.17 & 90.23 & 89.50 \\
Small-scale SFT + RL & \textbf{61.20} & \textbf{63.50} & \textbf{62.15} & \textbf{93.68} & 90.23 & \textbf{92.62} & \textbf{92.01} \\
\bottomrule
\end{tabular}}
\end{table}

\subsubsection{\textbf{Implementation Details.}}
We employ Qwen3.5-35B~\cite{qwen3.5} as the student backbone for both SFT and RL, and adopt the larger Qwen3.5-397B~\cite{qwen3.5} as the teacher for Stage-2 distillation. We use GLM-4.7~\cite{5team2025glm45agenticreasoningcoding} for prompt optimization, SFT data quality filtering, and thinking-quality evaluation, while DeepSeek-V4-Flash~\cite{deepseekai2026deepseekv4} serves as the reward model during RL, thereby separating reward modeling from evaluation to mitigate reward hacking. We set the maximum context length $L_{\max}$ to 32K tokens, and set $q_{\mathrm{ad}}$, $q_{\mathrm{content}}$, and $q_{\mathrm{profile}}$ to 20K, 7K, and 5K tokens, respectively. The quality threshold $\eta_{\mathrm{sft}}$ is set to 80 on a 0--100 scale. For the rubric-based thinking reward, the complete weight settings are reported in Appendix~\ref{appendix:reward_weights}. Detailed training hyperparameters are deferred to Appendix~\ref{appendix:more_details}.

\subsection{Overall Results~(RQ1)}
We compare DASH with prompting-based methods on diverse LLMs. We report three variants of DASH: 
``Small-scale SFT'' and ``Large-scale SFT'' are trained on 4K and 12K quality-filtered distilled samples, respectively, while ``Small-scale SFT + RL'' further applies hybrid-reward GRPO initialized from ``Small-scale SFT''. 
From the results in Table~\ref{tab:result_all}, we draw three main observations:

\begin{itemize}[leftmargin=1em, itemsep=0pt, topsep=0pt, parsep=0pt, partopsep=0pt]
    \item DASH with ``Small-scale SFT + RL'' achieves the best action and thinking performance in most cases. It consistently outperforms strong general-purpose models despite using the substantially smaller student backbone, and also achieves clear improvements over the original student model in both action prediction and thinking quality. 
    The improvement arises from the complementary roles of the three stages: CE extracts decision-relevant signals from heterogeneous histories, SFT equips the student model with basic simulation capabilities through trajectory distillation, and RL jointly aligns action prediction and thinking quality.
    \item 
    Despite that DASH trained with SFT (both small-scale and large-scale data) yields moderate performance gains in action quality, the thinking quality remains limited compared to the baselines. 
    This indicates that simulation depends not only on model and data scale, but also on effective domain adaptation and objective alignment (\ie thinking reward). 
    Further analysis on the effect of SFT data scale is discussed in Section~\ref{sec:sft_analysis}. 
    \item An interesting observation is that logic quality tends to be more closely associated with action prediction than content quality. ``Small-scale SFT + RL'' achieves both the highest logic score and the highest Weighted-F1, whereas models with strong content quality do not necessarily achieve accurate action prediction. For example, MiniMax-M2.5 obtains a relatively high content score but the lowest Weighted-F1. This suggests that advertising simulation depends not only on generating factually plausible thinking, but also on maintaining causal coherence and deriving the final action consistently from the identified signals.

\end{itemize}

\subsection{In-depth Analysis}
\label{sec:analysis}

\subsubsection{\textbf{Ablation Study~(RQ2)}}
\begin{table}[t]
  \centering
  \captionsetup{belowskip=0pt, aboveskip=0pt}
  \caption{Ablation study of the key components in DASH.}
  \label{tab:component_ablation}
  \setlength{\tabcolsep}{9pt}
  \setlength{\extrarowheight}{1pt}
  \setlength{\aboverulesep}{0pt}
  \setlength{\belowrulesep}{0pt}
  \small
  \begin{tabular}{lcccc}
    \toprule
    \textbf{Method} & \textbf{W-Prec.} & \textbf{W-Rec.} & \textbf{W-F1} & \textbf{Think.} \\
    \midrule
    \rowcolor{gray!15}
    \textbf{DASH} & \textbf{61.20} & \textbf{63.50} & \textbf{62.15} & \textbf{92.01} \\
    \quad w/o CE  & 54.69 & 52.60 & 51.93 & 89.20 \\
    \quad w/o SFT & 59.08 & 61.30 & 59.94 & 90.82 \\
    \quad w/o RL  & 59.69 & 61.30 & 59.07 & 90.10 \\
    \bottomrule
  \end{tabular}
\end{table}
To investigate the contribution of each key component, we conduct an ablation study by separately removing CE, SFT, and RL from DASH. Concretely, the variant without CE removes the CE stage while retaining SFT and RL; the variant without SFT directly applies RL to the base model; and the variant without RL stops after ``Small-scale SFT''.

As shown in Table~\ref{tab:component_ablation}, 
1) CE has the largest impact on overall performance, indicating that effective context construction is the foundation of accurate user simulation. 
2) RL contributes more directly to the final performance gains than SFT, highlighting the effectiveness of the hybrid reward in jointly optimizing action prediction and thinking quality. Nevertheless, 
3) applying RL without SFT remains inferior to the complete pipeline, suggesting that SFT, although contributing a smaller standalone gain, provides an important task-adapted initialization that enables more effective policy optimization. More detailed analyses of each component are provided in the following subsections.

\subsubsection{\textbf{Effect of Cross-domain Histories~(RQ3).}}
\begin{table}[t]
  \centering
  \captionsetup{belowskip=0pt, aboveskip=0pt}
  \caption{Effect of cross-domain history.}
  \label{tab:ablation_domain}
  \setlength{\tabcolsep}{5pt}
  \setlength{\extrarowheight}{1pt}
  \setlength{\aboverulesep}{0pt}
  \setlength{\belowrulesep}{0pt}
  \small
  \begin{tabular}{cc ccc c}
    \toprule
    \textbf{Ad} & \textbf{Content} & \textbf{W-Prec.} & \textbf{W-Rec.} & \textbf{W-F1} & \textbf{Think.} \\
    \midrule
    \ding{55} & \ding{55} & 36.46 & 41.90 & 37.83 & 85.68 \\
    \ding{55} & \ding{51} & 36.09 & 41.60 & 37.42 & 86.92 \\
    \ding{51} & \ding{55} & 58.80 & 60.70 & 58.88 & 90.41 \\
    \ding{51} & \ding{51} & \textbf{59.57} & \textbf{61.30} & \textbf{60.23} & \textbf{91.43} \\
    \bottomrule
  \end{tabular}
\end{table}
To investigate the impact of cross-domain history on user simulation, we conduct an ablation study over the two historical streams used in our compressed context, namely advertising history $\mathcal{H}_u^{\text{ad}}$ and cross-domain content history $\mathcal{H}_u^{\text{content}}$, using Qwen3.5-397B for experiments. We compare four configurations that progressively introduce each stream on top of the no-history baseline, while keeping the prompt template and context budget fixed across runs. 

As shown in Table~\ref{tab:ablation_domain}, we find that:
1) Combining advertising and content histories achieves the best overall performance. This confirms our claim in the Section~\ref{sec:intro} that users' responses to a target ad are jointly shaped by heterogeneous behavioral signals. Advertising history provides direct evidence of ad-specific preferences and conversions, while content history captures users’ evolving interests and current intent.
2) Advertising history alone substantially improves both action prediction and thinking quality, while adding content history yields further gains. 
However, 3) using content history alone causes a performance drop in action prediction because general content interests are not directly aligned with advertising responses. Thus, content history is most effective as a complementary signal when combined with advertising history.

\subsubsection{\textbf{Effect of Prompt Optimization~(RQ3).}} To assess the effect of our closed-loop prompt optimization procedure, we compare two prompts on the Qwen3.5-397B with identical input context and evaluation pipeline: (i) a \emph{Vanilla} prompt that gives only a minimal task instruction (Appendix~\ref{appendix:prompt_vanilla}), and (ii) the \emph{Optimized} prompt produced by our closed-loop optimizer~(Appendix~\ref{appendix:prompt_ours}).

\begin{figure}[t]
  \centering
  \captionsetup{belowskip=0pt, aboveskip=0pt}
  \includegraphics[width=0.85\columnwidth]{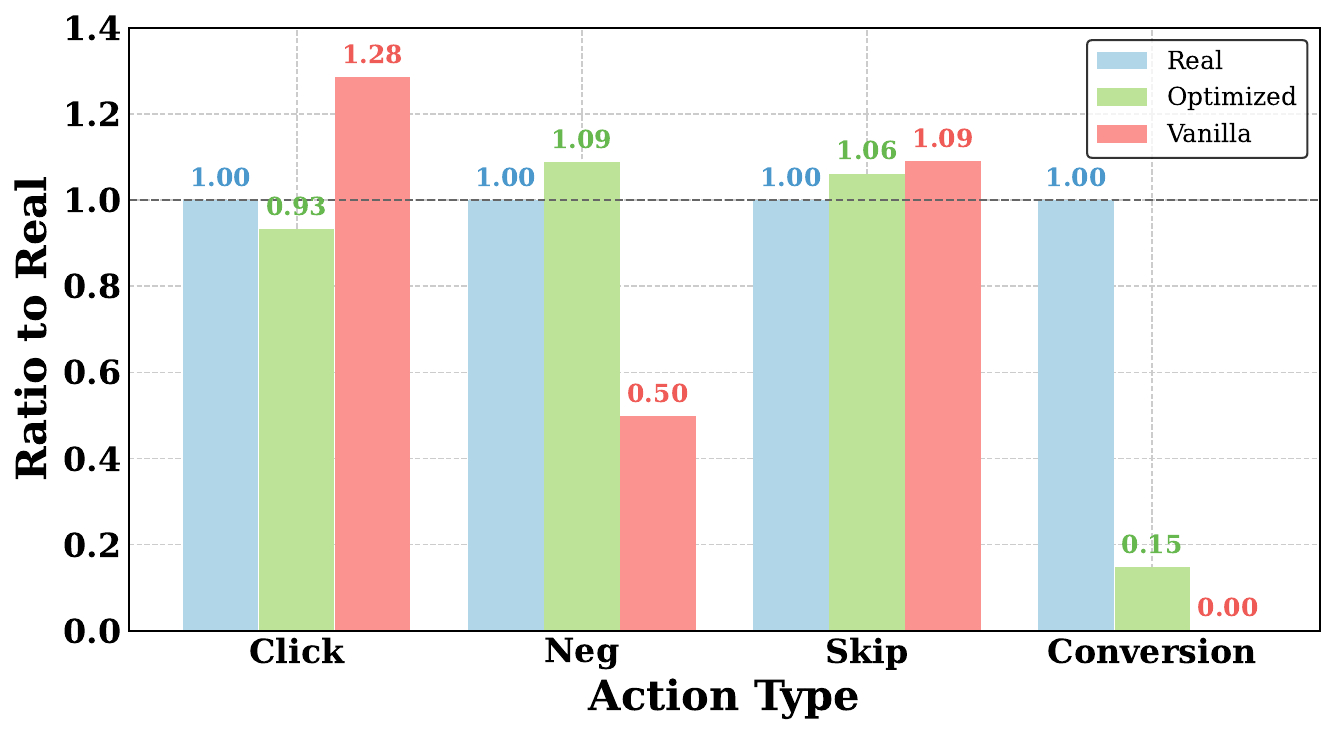}
  \caption{Ratio of predicted to real action frequency under Vanilla and Optimized prompts.}
  \label{fig:prompt_design}
\end{figure}

We make two observations: 1) As shown in Table~\ref{tab:ablation_prompt}, the optimized prompt substantially improves all action-prediction metrics, as well as the Content and Logic dimensions of thinking quality, while causing only a modest decrease in Form. This decrease may arise because prompt optimization focuses on recurring errors, such as unsupported claims and inconsistencies between the thinking trace and the predicted action. The resulting grounding and consistency constraints may slightly reduce the naturalness and diversity of the generated traces. 
2) From the action distribution perspective, Figure~\ref{fig:prompt_design} shows that the optimized prompt produces predictions that are more closely aligned with the empirical action distribution. In particular, it reduces the vanilla prompt's tendency to over-predict clicks and under-predict negative feedback and conversions.

\begin{table}[t]
  \centering
  \captionsetup{belowskip=0pt, aboveskip=0pt}
  \caption{Effect of prompt design.}
  \label{tab:ablation_prompt}
  \setlength{\tabcolsep}{5pt}
  \setlength{\extrarowheight}{1pt}
  \setlength{\aboverulesep}{0pt}
  \setlength{\belowrulesep}{0pt}
  \small
  \begin{tabular}{l ccc ccc}
    \toprule
    \textbf{Prompt} & \textbf{W-Prec.} & \textbf{W-Rec.} & \textbf{W-F1} & \textbf{Form} & \textbf{Content} & \textbf{Logic} \\
    \midrule
    Vanilla & 53.40 & 51.10 & 50.47 & \textbf{94.57} & 88.29 & 89.10 \\
    Optimized & \textbf{59.57} & \textbf{61.30} & \textbf{60.23} & 92.96 & \textbf{93.28} & \textbf{90.25} \\
    \bottomrule
  \end{tabular}
\end{table}

\begin{figure*}[t]
\centering
\captionsetup{belowskip=0pt, aboveskip=0pt}
\includegraphics[width=0.95\textwidth]{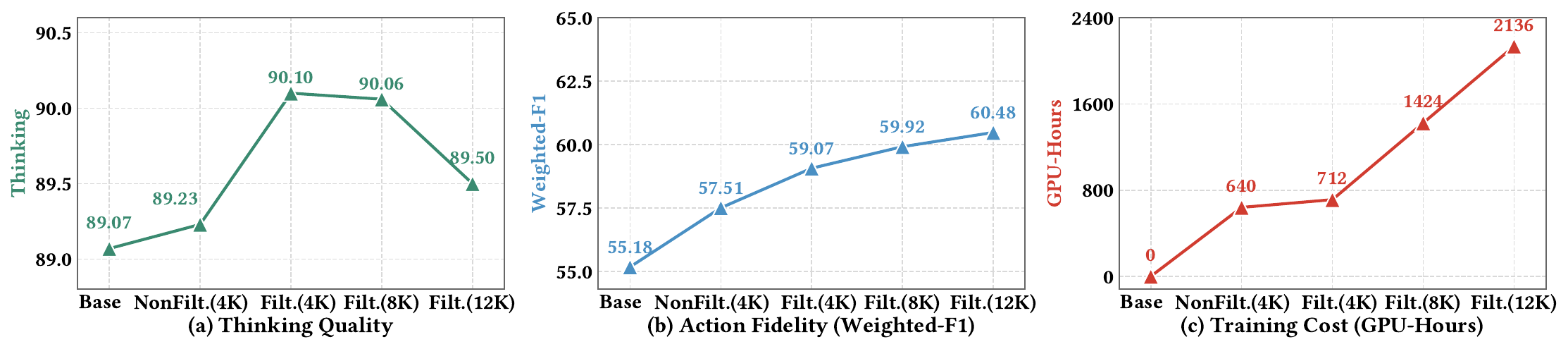}
\caption{SFT ablation across data quality and scale. Base: no task-specific SFT; NonFilt.(4K): 4K unfiltered trajectories; Filt.(4K/8K/12K): 4K, 8K, or 12K quality-filtered trajectories.}
\label{fig:sft_variants}
\end{figure*}

\subsubsection{\textbf{Effect of SFT Data Quality and Scale~(RQ4).}} 
\label{sec:sft_analysis}
We further analyze the effect of SFT by comparing five variants with increasing SFT data size and quality. Specifically, \textit{Base} denotes the student model without task-specific fine-tuning; \textit{NonFilt.(4K)} applies SFT using 4K teacher-distilled trajectories without thinking-quality filtering; \textit{Filt.(4K)} uses 4K trajectories following our data curation process (\cf \ref{sec:data_curation}); and \textit{Filt.(8K)} and \textit{Filt.(12K)} scale the data size to 8K and 12K high-quality samples, respectively. 
To quantify the computational cost of SFT, we report the training cost in GPU-hours, including data curation and student-model fine-tuning.

As shown in Figure~\ref{fig:sft_variants}, we make two observations: 
1) SFT effectively adapts the student model to the advertising user simulation task, substantially improving Weighted-F1. This result demonstrates that teacher-distilled trajectories provide effective supervision for learning task-specific action patterns, whereas the improvement in thinking quality remains marginal. 
2) Increasing the scale of the distilled training data yields only limited additional gains, while thinking quality begins to decrease and the end-to-end training cost increases substantially. These observations indicate that simply scaling SFT is an inefficient approach to continued improvement. This limitation stems from SFT being constrained by the coverage of distilled trajectories. Balancing simulation performance and training cost, we therefore adopt Filt. (4K) as our ``Small-scale SFT'' configuration, which provides a cost-effective initialization for RL to further improve performance through on-policy exploration.

\subsubsection{\textbf{Effect of Reward Design and Initialization~(RQ4).}} 
\label{sec:rl_analysis}
\begin{table}[t]
  \centering
  \captionsetup{belowskip=0pt, aboveskip=0pt}
  \caption{Effect of RL reward composition and initialization.}
  \label{tab:ablation_rl}
  \setlength{\tabcolsep}{6pt}
  \setlength{\extrarowheight}{1pt}
  \setlength{\aboverulesep}{0pt}
  \setlength{\belowrulesep}{0pt}
  \small
  \begin{tabular}{ll cc}
    \toprule
    \textbf{Initialization} & \textbf{Reward} &
    \textbf{Weighted-F1} & \textbf{Think.}\\
    \midrule
    \rowcolor{gray!15}
    Base & -- & 55.18 & 89.07 \\
     & Action & 59.83 & 87.26 \\
     & Action+Think & 59.94 & 90.82 \\
    \midrule
    \rowcolor{gray!15}
    Small-scale SFT & -- & 59.07 & 90.10 \\
     & Action & 60.56 & 88.92 \\
     & Action+Think &
    \textbf{62.15} & \textbf{92.01} \\
    \bottomrule
  \end{tabular}
\end{table}
We further dissect how the GRPO reward composition and the choice of starting checkpoint affect the simulator's action prediction and thinking quality. We compare four RL variants that toggle the inclusion of the rubric-based thinking reward and the use of a ``Small-scale SFT'' warm-up before RL.

As shown in Table~\ref{tab:ablation_rl}, we make three observations. 
1) SFT provides RL with a task-adapted initialization, establishing basic capabilities for both action prediction and thinking generation before policy optimization. Starting from the Base checkpoint, action-only RL improves action prediction but reduces thinking quality. This pattern is consistent with the decrease in thinking quality observed in Section~\ref{sec:sft_analysis}, suggesting that optimizing only outcome-level action correctness may weaken the quality of the underlying thinking process. 
2) Introducing the rubric-based thinking reward addresses the above limitation and improves both action prediction and thinking quality by providing dense supervision for thinking traces. 
3) Combining the SFT initialization with the hybrid reward yields the strongest overall performance. This combination reflects their complementary roles: SFT equips the model with task-specific simulation capabilities, while the hybrid reward further aligns the thinking traces with the predicted action.

\begin{figure}[t]
  \centering
  \captionsetup{belowskip=0pt, aboveskip=0pt}
  \includegraphics[width=0.9\columnwidth]{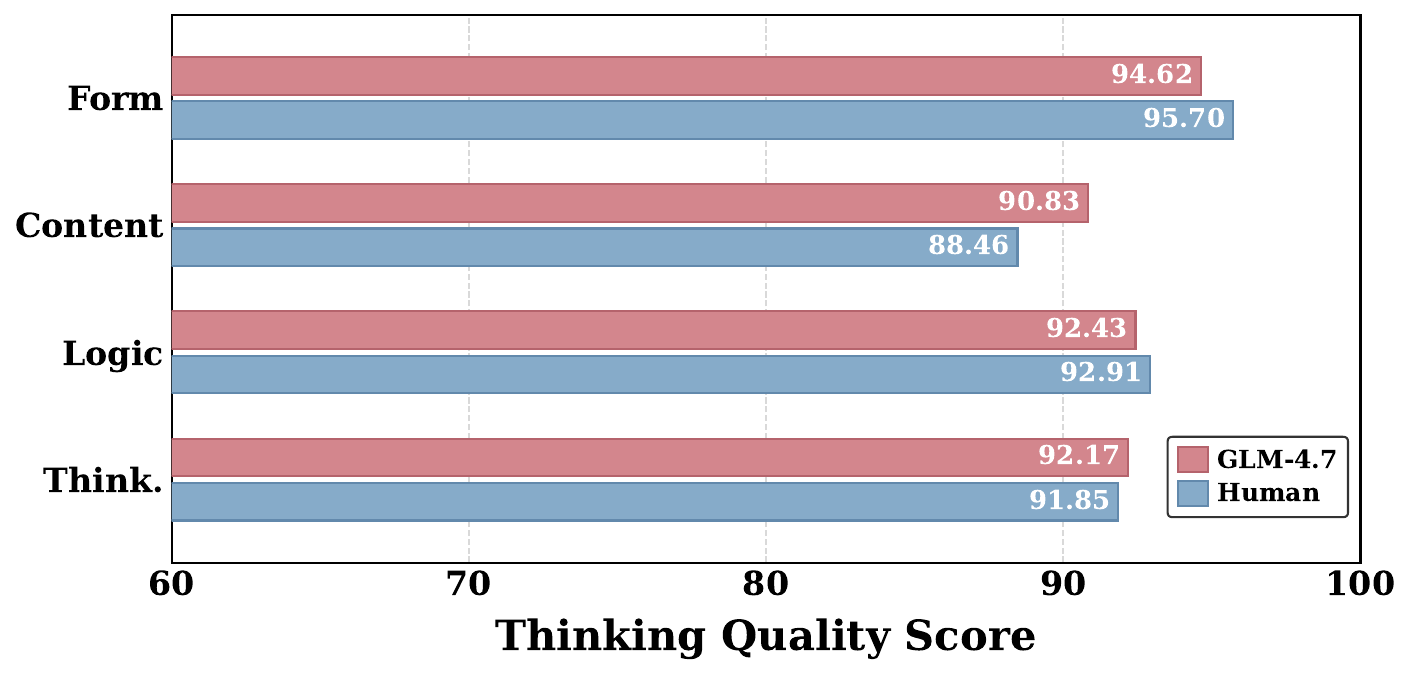}
  \caption{Alignment between LLM-based evaluator scores and human judgments across thinking-quality dimensions.}
  \label{fig:llm_align}
\end{figure}

\subsubsection{\textbf{Alignment with Human Judgements~(RQ5).}}

To validate whether the LLM-based evaluator produces judgments consistent with human assessment, we randomly sample 100 outputs generated by DASH from the test set. Three human annotators independently evaluate the same set of outputs using the same evaluation rubrics as GLM-4.7, covering the three dimensions of Form, Content, and Logic. We average the scores assigned by the three annotators to obtain the human evaluation score for each dimension.

As shown in Figure~\ref{fig:llm_align}, the scores produced by GLM-4.7 are highly consistent with those assigned by human annotators across Form, Content, Logic, and overall thinking quality. This similarity indicates that the LLM-based evaluator follows evaluation standards that are broadly aligned with human judgment and can capture the quality differences among thinking traces. The agreement also suggests that the LLM-based evaluator can provide reliable results while reducing the cost of human annotation. 

\subsubsection{\textbf{More Analysis}}
Due to space constraints, we provide additional analyses of action prediction across head and long-tail actions in Appendix~\ref{appendix:action_breakdown_analysis}, the cross-evaluator robustness of thinking-quality assessment in Appendix~\ref{appendix:robustness}, and three representative case studies that show interpretable diagnostic value in Appendix~\ref{appendix:case_study}.

\section{Conclusion and Future Work}
\label{sec:conclusion}

In this work, we presented DASH, an LLM-based user behavior simulator for industrial advertising. Departing from conventional simulators that merely reproduce observable action labels, DASH jointly models action prediction and thinking traces, supported by a dual-metric evaluation protocol that assesses both action fidelity and thinking quality along Form, Content, and Logic dimensions. 

Looking ahead, we are actively exploring how user simulators can be integrated into real advertising systems as practical infrastructure. We envision three directions: 1) offline A/B test pre-screening, where the simulator serves as a proxy to filter candidate strategies before live experimentation; 
2) pre-test for ad recommendation, where the simulator evaluates candidate ads before they are served to users, helping identify ineffective or potentially harmful ads and optimize ad selection and recommendation strategies; 
and 3) recommendation optimization, where exposing thinking traces provides diagnostic signals for refining ad creatives and balancing monetization with user experience. 
More broadly, we envision DASH evolving into a controllable advertising sandbox for accelerating the iteration of safer, more user-centered advertising systems.


\section*{Ethics and Privacy Statement}

This work studies an LLM-based simulator of advertising user behavior built on top of cross-domain interaction logs collected from a real-world advertising platform. All training and evaluation data are processed on internal infrastructure under the platform's standard privacy and data-governance policies: user identifiers are hashed, raw content is restricted to internally accessible features, and no personally identifiable information is included in the released artifacts. We acknowledge that any high-fidelity user simulator carries dual-use risks, including the potential to fit overly aggressive ranking or targeting strategies on simulated users; to mitigate this, we restrict DASH to evaluative, diagnostic, and augmentation use cases. The dataset, prompts, and evaluation rubrics described in this paper are used solely for research purposes and are reported in aggregate.

\bibliographystyle{ACM-Reference-Format}
\bibliography{ref}

\appendix
\section{Dataset Details}
\label{appendix:dataset_details}

\subsection{Data Collection}
\label{data_collection}
We organize data collection on the Tencent platform at the granularity of individual ad-request interactions. For each ad request served to user $u$ at timestamp $t$, the corresponding sample is assembled from the following five components. The two behavior sequences ($\mathcal{H}_u^{\text{ad}}$ and $\mathcal{H}_u^{\text{content}}$) are constructed by retrieving \emph{all} of the user's interactions within the 30-day window strictly preceding $t$, ensuring temporal causality and preventing label leakage:

\begin{itemize}[leftmargin=1em, itemsep=0pt, topsep=0pt]
  \item \textbf{Ad Behavior Sequence} ($\mathcal{H}_u^{\text{ad}}$): The user's chronologically ordered advertising interaction history spanning the full platform, recording skips, clicks, conversions, and negative-feedback regardless of the serving domain.
  \item \textbf{Content Behavior Sequence} ($\mathcal{H}_u^{\text{content}}$): The user's native content browsing behaviors drawn from five heterogeneous domains (App, Game, Video, Article, and Search), capturing the organic patterns that shape cognitive state and preferences.
  \item \textbf{User Profile} ($p_u$): Persistent user attributes including demographics, consumption habits, device information, and longitudinal interest profiles.
  \item \textbf{Ad Profile} ($d_t$): Creative attributes of the candidate advertisement, including category, industry tag, copywriting text, OCR-parsed image content, product pricing, and promotion mechanism.
  \item \textbf{Ad Request Context} ($c_t$): The decision context at the moment of ad serving, including domain identifier, ad-slot position, spatio-temporal signals (timestamp, day-of-week), and geographic location.
\end{itemize}

\subsection{Content Behavior domain Descriptions}

The five content behavior domains span diverse user intent patterns and consumption modalities:

\begin{itemize}[leftmargin=1em, itemsep=0pt, topsep=0pt]
  \item \textbf{App}: A task-driven application domain covering utility, commerce, and lifestyle services. User behaviors include launching, browsing, and task completion within applications.
  \item \textbf{Game}: A gaming domain characterized by high-frequency, short-session engagement. User behaviors reflect gaming preferences, session patterns, and in-game interaction styles.
  \item \textbf{Video}: A short-form and medium-form video feed domain. User behaviors capture video viewing, liking, and sharing patterns that reveal topical interests and attention dynamics.
  \item \textbf{Article}: A long-form article-feed domain. User behaviors reflect subscription choices, reading depth, and content engagement patterns indicative of more deliberate information consumption.
  \item \textbf{Search}: An information retrieval domain driven by explicit user intent. User behaviors include search queries, click-through patterns, and result interactions that directly reveal intent signals.
\end{itemize}

\subsection{Dataset Statistics}
\begin{table}[h]
\centering
\small
\captionsetup{belowskip=0pt, aboveskip=0pt}
\caption{Per-request event and token statistics on the advertising dataset.}
\label{dataset_statistic}
\resizebox{\columnwidth}{!}{%
\begin{tabular}{lrrr}
\toprule
Domain & Avg.\ \#events / req. & Mean tokens / event & Tokens / ad request \\
\midrule
Ad domain      &   1,354.8 & 2722.1 & 3,688,008 \\
Content domain &     396.2 &   97.5 &    38,612 \\
\midrule
Combined       &   1,751.0 & --     & 3,726,620 \\
\bottomrule
\end{tabular}}
\end{table}


Table~\ref{dataset_statistic} reports the dataset statistics used to characterise our representation. To obtain a tractable summary of the production data distribution, we uniformly draw a random sample of $10{,}000$ ad requests, each originating from a distinct user, and treat this sample as a representative slice of the underlying ad-serving stream; all statistics reported in this section are computed on that sample.

The sampled ad requests exhibit a moderately imbalanced action distribution: skip accounts for approximately 45\% of the requests, followed by click at 35\%, negative-feedback at 15\%, and conversion at 5\%. This distribution highlights the long-tail characteristics of high-intent actions and explicit user feedback in real-world advertising traffic.

Beyond the action-label distribution, Table~\ref{dataset_statistic} also characterises the scale and granularity of the historical context associated with each request. On average, each ad request is accompanied by $1{,}354.8$ ad-domain interaction and $396.2$ content-domain interaction collected across the five Tencent content domains, yielding a combined behavioural history of approximately $1{,}751$ interaction per request. All historical interaction are drawn from the 30-day window strictly preceding the corresponding request timestamp. The two domains also differ substantially in interaction granularity, with nearly two orders of magnitude separating their average token counts per interaction ($2{,}722.1$ versus $97.5$). This asymmetry is intentional: ad-domain interaction provide direct evidence of users' prior responses to advertising and therefore retain detailed information, including the complete creative profile (\eg category, industry tag, and copywriting), OCR- and caption-derived multimodal descriptions of images or videos, product pricing, and other relevant attributes. Content-domain interaction, by contrast, serve primarily as background context for characterising the user's latent interest state and are consequently represented as compact metadata records containing the domain identifier, a short title or category, and a timestamp. Taken together, these statistics correspond to approximately $3.69$M raw tokens per request from the ad domain and $38.6$K from the content domain, yielding a combined raw context of about $3.73$M tokens per request.



\section{Reward and Evaluation Score Specification}
\label{appendix:reward_weights}

This appendix specifies the rubric definitions and score aggregation used for both the thinking-quality reward during RL and the thinking-quality evaluation. The evaluation framework decomposes thinking quality into three complementary dimensions: form, content, and logic, which are further divided into seven sub-dimensions and their corresponding fine-grained criteria. We first describe the three dimensions and their sub-dimensions, and then present the normalization, weighting, and hierarchical aggregation procedures used to compute $R_{\mathrm{think}}$ and the reported evaluation scores.

\begin{itemize}[leftmargin=1em, itemsep=0pt, topsep=0pt, parsep=0pt, partopsep=0pt]
    \item \textbf{Form}
    assesses how clearly and appropriately the thinking trace is
    expressed and organized through two sub-dimensions:
    \textbf{Linguistic Expression}, which evaluates
    Grammar Correctness, Fluency, Naturalness, and Diversity; and
    \textbf{Thinking-step Completeness}, which
    measures the coverage of decision-relevant thinking steps using the
    Thinking-step Coverage Score, inspired by the
    Engel--Blackwell--Miniard consumer decision-process
    model~\cite{engel1995consumer}.

    \item \textbf{Content}
    assesses whether the thinking trace is factually grounded in the
    observable decision context through two sub-dimensions:
    \textbf{Factual Correctness}, which evaluates
    Hallucination Rate and Overall Factual Accuracy; and
    \textbf{Business-knowledge Alignment}, which
    evaluates User Profile Alignment, Ad Attribute Alignment, and
    Scenario Context Alignment.

    \item \textbf{Logic}
    assesses whether the identified evidence forms a coherent decision
    process and supports the predicted action through three
    sub-dimensions:
    \textbf{Attribution Clarity}, which evaluates
    Key Factor Coverage and Evidence Support;
    \textbf{Thinking Coherence}, which evaluates
    Causal Coherence and Internal Consistency; and
    \textbf{Thinking--Action Alignment}, which
    evaluates Behavior Alignment between the generated thinking trace
    and the predicted action.
\end{itemize}

Given the above rubric hierarchy, the scores are aggregated using a three-level bottom-up scheme. First, individual rubric scores are normalized to $[0,1]$: LLM-assessed Likert scores are mapped via $\hat{s}=(s-1)/4$ for $s\in\{1,2,3,4,5\}$, while binary indicators are used directly. The normalized rubric scores are then aggregated into sub-dimension scores, dimension-level scores, and finally the overall thinking-quality reward $R_{\mathrm{think}}$. 


\paragraph{Form.}
\begin{align}
  R_{\text{exp}} &= \alpha_1 \, R_{\text{gram}} + \alpha_2 \, R_{\text{flu}} + \alpha_3 \, R_{\text{nat}} + \alpha_4 \, R_{\text{div}}, \\
  R_{\text{TC}} &= \frac{1}{7}\sum_{k=1}^{7} [\text{step } k \text{ present}].
\end{align}

\paragraph{Content.}
\begin{align}
  R_{\text{fact}} &= \beta_1 \, R_{\text{halluc}} + \beta_2 \, R_{\text{accuracy}}, \\
  R_{\text{align}} &= \beta_3 \, R_{\text{user}} + \beta_4 \, R_{\text{ad}} + \beta_5 \, R_{\text{ctx}}.
\end{align}

\paragraph{Logic.}
\begin{align}
  R_{\text{attr}} &= \gamma_1 \, R_{\text{keyfactor}} + \gamma_2 \, R_{\text{evidence}}, \\
  R_{\text{coher}} &= \gamma_3 \, R_{\text{causal}} + \gamma_4 \, R_{\text{consist}}, \\
  R_{\text{behav}} &= \gamma_5 \,R_{\text{ba}}.
\end{align}

\textbf{Dimension weights.}
At the dimension level, we set $w_1=0.10$ for Form, $w_2=0.30$ for Content, and $w_3=0.60$ for Logic.

\textbf{Sub-dimension weights.}
At the sub-dimension level, we set $a_1=0.70$ for Linguistic Expression and $a_2=0.30$ for Thinking-step Completeness within Form;
$b_1=0.60$ for Factual Correctness and $b_2=0.40$ for Business-knowledge Alignment within Content;
and $c_1=0.35$ for Attribution Clarity, $c_2=0.35$ for Thinking Coherence, and $c_3=0.30$ for Thinking--Action Alignment within Logic.

\textbf{Rubric weights.}
At the rubric level, we set $\alpha_1=0.25$ for Grammar Correctness, $\alpha_2=0.30$ for Fluency, $\alpha_3=0.25$ for Naturalness, and $\alpha_4=0.20$ for Diversity within Linguistic Expression;
$\beta_1=0.50$ for Hallucination Rate and $\beta_2=0.50$ for Overall Factual Accuracy within Factual Correctness;
$\beta_3=1/3$ for User Profile Alignment, $\beta_4=1/3$ for Ad Attribute Alignment, and $\beta_5=1/3$ for Scenario Context Alignment within Business-knowledge Alignment;
$\gamma_1=0.50$ for Key Factor Coverage and $\gamma_2=0.50$ for Evidence Support within Attribution Clarity;
$\gamma_3=0.50$ for Causal Coherence and $\gamma_4=0.50$ for Internal Consistency within Thinking Coherence;
and $\gamma_5=1.00$ for Behavior Alignment within Thinking--Action Alignment.

Notably, in our advertising application, we place greater emphasis on the logic dimension, as causal coherence and thinking–action consistency are particularly important for reliable user simulation. Accordingly, we assign a larger weight to this dimension. More generally, we recommend adapting the dimension weights to the priorities of each target application. For reporting purposes, all thinking quality numbers in the experimental tables are presented on a 0--100 scale obtained by $R_{\text{think}} \times 100$.

\section{Additional Experimental Results}
\label{appendix:additional_results}

\subsection{Head and Long-tail Action Analysis}
\label{appendix:action_breakdown_analysis}

Aggregate weighted metrics may obscure whether the improvements are concentrated on frequent behaviors or extend to sparse but business-critical actions. We therefore partition the action space into \textbf{head actions} (skip and click) and \textbf{long-tail actions} (conversion and negative-feedback), and report support-weighted precision, recall, and F1 within each group in Table~\ref{tab:action_breakdown}.

``Small-scale SFT + RL'' achieves the strongest performance on both head and long-tail actions. Its gains are not limited to dominant behaviors; the proposed training pipeline also improves the recognition of sparse, high-information responses, indicating a more balanced modeling of the overall action distribution.

The stage-wise comparison further reveals the complementary roles of SFT and RL. Small-scale SFT improves the model's sensitivity to long-tail actions, suggesting that hard samples provide useful supervision for behavioral patterns that are underrepresented in ordinary imitation data. However, this increased sensitivity also introduces more false-positive predictions. Building on the SFT initialization, RL with the hybrid action--thinking reward improves the discrimination of both head and long-tail actions. These results suggest that rubric-based reasoning supervision does not merely encourage the model to predict rare actions more frequently; instead, it helps the model determine whether the available signal genuinely supports the predicted action.

Compared with the substantially larger models, they exhibit higher long-tail precision but lower recall, whereas our model retrieves a larger fraction of sparse behaviors. This different operating point is particularly desirable for simulation-oriented applications, where systematically missing conversions or negative-feedback interactions can distort offline strategy evaluation. ``Large-scale SFT'' obtains a marginally higher long-tail precision than ``Small-scale SFT + RL'', but the latter achieves higher long-tail recall and substantially stronger head-action performance. Overall, ``Small-scale'' SFT followed by hybrid-reward RL provides the most balanced operating point across behavior frequencies.

\begin{table}[t]
\centering
\captionsetup{belowskip=0pt, aboveskip=0pt}
\caption{Per-group action-quality breakdown.}
\label{tab:action_breakdown}
\setlength{\tabcolsep}{3pt}
\setlength{\extrarowheight}{0pt}
\renewcommand{\arraystretch}{0.95}
\setlength{\aboverulesep}{0pt}
\setlength{\belowrulesep}{0pt}
\small
\begin{tabular}{l ccc ccc}
\toprule
\multirow{2}{*}{\textbf{Model}}
 & \multicolumn{3}{c}{\textbf{Head} ($\uparrow$)}
 & \multicolumn{3}{c}{\textbf{Long-tail} ($\uparrow$)} \\
\cmidrule(lr){2-4}\cmidrule(lr){5-7}
 & W-P & W-R & W-F1
 & W-P & W-R & W-F1 \\
\midrule
Qwen3.5-35B        & 54.96 & 53.35 & 54.12 & 52.25 & 66.53 & 58.53 \\
DeepSeek-V4-Flash  & 53.29 & \underline{60.71} & 56.72 & 67.72 & 43.93 & 53.29 \\
MiniMax-M2.5       & 47.28 & 56.58 & 50.76 & \underline{72.18} & 28.87 & 41.24 \\
Qwen3.5-397B       & 58.67 & 59.66 & \underline{59.00} & \textbf{74.99} & 66.52 & 64.14 \\
Kimi-K2.5          & 56.46 & 54.27 & 54.91 & 61.43 & 69.04 & 61.42 \\
\midrule
\rowcolor{black!7}\multicolumn{7}{l}{\textbf{DASH}~(Qwen3.5-35B)} \\
Small-scale SFT    & \underline{60.78} & 58.08 & 57.86 & 56.20 & \underline{70.54} & 62.95 \\
Large-scale SFT    & 58.64 & 59.26 & 58.87 & 62.18 & 69.46 & \underline{65.39} \\
Small-scale SFT + RL & \textbf{61.41} & \textbf{61.10} & \textbf{61.13} & 60.51 & \textbf{71.13} & \textbf{65.62} \\
\bottomrule
\end{tabular}
\end{table}

\subsection{Evaluator Robustness}
\label{appendix:robustness}
To assess evaluator robustness, we select Qwen3.5-397B and Qwen3.5-35B, two models from the same family with a known performance gap~\cite{qwen3.5}, and independently evaluate their outputs using GLM-4.7~\cite{5team2025glm45agenticreasoningcoding}, DeepSeek-V4-Flash~\cite{deepseekai2026deepseekv4}, and Hy3-preview~\cite{hunyuan3preview}. 

Table~\ref{tab:evaluator_robustness} shows that the comparative conclusions are insensitive to the choice of evaluator. Across the three evaluators, the overall Thinking scores vary by only 1.18 points for Qwen3.5-397B and 1.60 points for Qwen3.5-35B, with coefficients of variation of 0.66\% and 0.91\%, respectively. More importantly, Qwen3.5-397B consistently outperforms Qwen3.5-35B under every evaluator and across all four quality dimensions, resulting in a \textbf{100\%} pairwise ranking-consistency rate over 12 comparisons. The overall Thinking-quality advantage is also stable, ranging from 2.36 to 2.78 points, with a mean gap of ${2.58 \pm 0.21}$. These results indicate that evaluator choice primarily introduces small calibration shifts without altering model rankings or the main comparative conclusion.

\begin{table}[t]
\centering
\captionsetup{belowskip=0pt, aboveskip=0pt}
\caption{Cross-evaluator robustness of Thinking Quality.}
\label{tab:evaluator_robustness}
\footnotesize
\setlength{\tabcolsep}{2.2pt}
\renewcommand{\arraystretch}{1.05}
\resizebox{\columnwidth}{!}{%
\begin{tabular}{llccccccc}
\toprule
\textbf{Model}
& \textbf{Evaluator}
& \textbf{Form}
& \textbf{Content}
& \textbf{Logic}
& \textbf{Think.}
& \shortstack{\textbf{Think.}\\\textbf{Mean$\pm$SD}}
& \shortstack{\textbf{Think.}\\\textbf{Range}}
& \textbf{CV (\%)} \\
\midrule

\multirow{3}{*}{Qwen3.5-397B}
  & GLM-4.7
  & 92.96 & 93.28 & 90.25 & 91.43
  & \multirow{3}{*}{$\mathbf{90.80\pm0.60}$}
  & \multirow{3}{*}{\textbf{1.18}}
  & \multirow{3}{*}{\textbf{0.66}} \\

  & DeepSeek-V4-Flash
  & 89.99 & 91.19 & 89.83 & 90.25
  & & & \\

  & Hy3-preview
  & 91.72 & 90.14 & 90.83 & 90.71
  & & & \\

\midrule

\multirow{3}{*}{Qwen3.5-35B}
  & GLM-4.7
  & 90.20 & 88.86 & 88.99 & 89.07
  & \multirow{3}{*}{$\mathbf{88.22\pm0.81}$}
  & \multirow{3}{*}{\textbf{1.60}}
  & \multirow{3}{*}{\textbf{0.91}} \\

  & DeepSeek-V4-Flash
  & 88.24 & 89.73 & 86.22 & 87.47
  & & & \\

  & Hy3-preview
  & 89.83 & 87.42 & 88.18 & 88.11
  & & & \\

\midrule

\multicolumn{2}{l}{\textbf{Cross-model Thinking gap}}
& \multicolumn{4}{c}{$\mathbf{2.58\pm0.21}$}
& \multicolumn{3}{c}{Range: $\mathbf{2.36\text{--}2.78}$} \\

\multicolumn{2}{l}{\textbf{Pairwise ranking consistency}}
& \multicolumn{7}{c}{
  $\mathbf{100\%}$ (12/12 evaluator--dimension comparisons)
} \\

\bottomrule
\end{tabular}%
}
\end{table}

\subsection{Case Study}
\label{appendix:case_study}

\definecolor{vanRed}{RGB}{200,60,60}
\definecolor{oursGreen}{RGB}{40,140,80}
\definecolor{verifyAmber}{RGB}{180,120,30}
\newtcolorbox{vanillabox}[1][]{%
  enhanced, breakable=true,
  colback=vanRed!4, colframe=vanRed!55,
  title=#1, fonttitle=\bfseries\small,
  fontupper=\footnotesize,
  left=4pt, right=4pt, top=3pt, bottom=3pt,
  arc=1mm, boxrule=0.5pt,
}
\newtcolorbox{oursbox}[1][]{%
  enhanced, breakable=true,
  colback=oursGreen!4, colframe=oursGreen!55,
  title=#1, fonttitle=\bfseries\small,
  fontupper=\footnotesize,
  left=4pt, right=4pt, top=3pt, bottom=3pt,
  arc=1mm, boxrule=0.5pt,
}

This appendix complements the quantitative gains reported in Section~\ref{sec:analysis} with three representative qualitative cases. The three cases target: (a) \emph{missing current-intent signal when the content-domain history is removed}, exposing the necessity of cross-domain content cues that are individually weak but jointly diagnostic of the user's current intent; (b) \emph{losing the most decisive signal buried inside a long behavior history}, a manifestation of the well-documented \emph{lost-in-the-middle} phenomenon~\citep{liu2024lost}; and (c) \emph{fabricating unobserved signal} to rationalize a high-engagement decision.

\smallskip
\noindent\textbf{(a) Missing current-intent signal without content-domain history.}\hspace{0.3em}
\emph{Setup.} We compare Qwen3.5-397B with the content-domain stream ablated (Ad-Only) against the same backbone with full ad-plus-content history, holding the prompt fixed (\ie rows 3 vs.\ 4 of Table~\ref{tab:ablation_domain}).

\smallskip
\noindent The user has a short-term interest tag pinned to furniture / storage cabinets; his ad-history is dominated by furniture and home-decor clips. Three independent cross-domain streams in the past week jointly indicate an active outbound-travel context that no single ad-history signal can reveal: (i) a same-day visit to the travel app plus a visit to the mobility app 7 days ago; (ii) a same-day full-completion of a \emph{safe-driving techniques} clip together with a \emph{seaside-garden travel} clip on Video; (iii) eight cross-day Search queries on outbound travel-route plans. The candidate ad is \textbf{Travel -- flight discount}, served at Saturday noon on the Video comment-area slot. Ground-truth action: \textbf{click}.

\vspace{0.3em}
\noindent
\begin{vanillabox}[Qwen3.5-397B (Ad-Only History) $\rightarrow$ \texttt{skip} \ding{55}]
\begin{CJK}{UTF8}{gbsn}
\textbf{\textit{Thinking.}} An ad for a travel service... I tend to be fairly cautious about spending. Most of the ads I have been seeing on social media lately are about furniture and home furnishings, and I have opened quite a few of them to learn more, so that is where my attention has been. As for travel, \textbf{my schedule is usually pretty fixed}, so I would normally plan a trip well in advance rather than buy a ticket at the last minute. The ad is also presented as a scripted conversation between a couple claiming a coupon, which I find a little annoying and not really in tune with my mood while I am quietly scrolling through videos... At the end of the day, it has nothing to do with the furniture I am currently interested in. It feels like a generic recommendation, so there is little reason to open it. I will just scroll past.\\[2pt]
\textbf{\textit{Judge.}} \emph{Cross\_domain\_unavailable}: with content-domain history removed, the model anchors on the dominant furniture ad stream and the long-term \emph{furniture} tag, and fabricates an ``unsupported schedule-based'' rationalization; the cues revealing the user's current travel intent reside entirely in the ablated stream and are unreachable.
\end{CJK}
\end{vanillabox}

\vspace{0.4em}
\noindent
\begin{oursbox}[Qwen3.5-397B (Ad + Content History) $\rightarrow$ \texttt{click} \ding{51}]
\begin{CJK}{UTF8}{gbsn}
\textit{Thinking.} A discounted-flight ad? This is actually well timed--I checked flights \textbf{in the Travel mini app just this morning}... \textbf{A couple of days ago, I also watched a short travel video featuring a seaside garden, and over the past few days I have searched for several driving routes for upcoming trips}. I have clearly had the idea of getting away on my mind lately. This flight ad appeared in the comments section of a video and highlights several promotional offers. It does not provide an exact price, but the travel service is a familiar brand and the platform seems reliable... That said, I have seen ads from other brands on social media without clicking them, so I am not sure whether this one will be any different. Still, I might actually be able to find a suitable ticket.

\textbf{\textit{Judge.}} \emph{cross\_domain\_grounded}: all claims input-locatable; the active travel context is \textbf{triangulated across App, Video, and Search} rather than the long-term furniture tag.
\end{CJK}
\end{oursbox}
\captionsetup{belowskip=0pt, aboveskip=0pt}
\captionof{figure}{Missing current-intent signal without content-domain history.}
\label{fig:case_a}

\smallskip
\noindent The decisive signal is the \emph{joint diagnosticity} of three individually weak cross-domain cues---an app touch, a seaside-garden travel video, and a week of outbound travel-route queries---all residing in the content-domain stream. Ablating that stream leaves only the furniture-dominated ad history and the static long-term tag, so the model anchors on dominant surface signal and fabricates an unsupported schedule-based rationalization. Restoring content history alone---without any change to the model or prompt---is sufficient for the same backbone to triangulate across App, Video, and Search and recover the correct \emph{click} decision, directly instantiating the ablation in Table~\ref{tab:ablation_domain}: the two streams are \emph{complementary in function}---ad history captures long-term preference, content history anchors current intent. Because these cues are low-volume relative to the dominant ad stream, DASH's stream-Level compression (Section~\ref{sec:compression}) reserves a dedicated content-stream budget so they survive compression in the million-token regime, and the \emph{Focus} phase (Section~\ref{sec:prompt_design}) retrieves them across all five content domains for downstream reasoning.

\smallskip
\noindent\textbf{(b) Ignoring strong historical negative signals.}\hspace{0.3em}
\emph{Setup.} On identical input, we compare an off-the-shelf Qwen3.5-397B with the vanilla prompt (Appendix~\ref{appendix:prompt_vanilla}) against the full DASH.

\smallskip
\noindent The user has a luxury-automotive long-term interest tag and mid-low consumption habit; the candidate ad is Luxury EV Model A on a social-feed slot at Saturday noon. The decisive signal in the input is two prior negative-feedback events on the same car (Luxury EV Model A, 7 days and 14 days before the request) and \emph{no} click or conversion on the brand anywhere in the 30-day history. Ground-truth action: \textbf{negative-feedback}.

\vspace{0.3em}
\noindent
\begin{vanillabox}[Qwen3.5-397B (Vanilla) $\rightarrow$ \texttt{click} \ding{55}]
\begin{CJK}{UTF8}{gbsn}
\textbf{\textit{Thinking.}} The dark-red backdrop and red interior make this car ad really eye-catching. As someone \textbf{who is interested in cars}, I \textbf{naturally} want to open it and check out the performance specifications and smart features... I have some free time around noon this weekend, so I will click through and take a look at the details, although I am not planning to buy one anytime soon.

\textbf{\textit{Judge.}} \emph{contradicted\_history (history\_intensity\_exaggeration)}: claims a click intent on a brand the user has \textbf{already negative-fed back twice in 14 days}; no click record exists. 
\end{CJK}
\end{vanillabox}

\vspace{0.4em}
\noindent
\begin{oursbox}[DASH~(Ours) $\rightarrow$ \texttt{negative-feedback} \ding{51}]
\begin{CJK}{UTF8}{gbsn}
\textbf{\textit{Thinking.}} Here we go again-another ad for this car. It has already been shown to me several times, and \textbf{I marked it as ``Not interested'' both seven and fourteen days ago...} I may have some interest in luxury cars, but a high-end electric vehicle like this does \textbf{not really fit} my spending habits... I will just mark it as ``Not interested'' again.\\[2pt]
\textbf{\textit{Judge.}} all claims grounded in input.
\end{CJK}
\end{oursbox}
\captionsetup{belowskip=0pt, aboveskip=0pt}
\captionof{figure}{Lost-in-the-middle on a strong negative signal.}
\label{fig:case_b}

\smallskip
\noindent The vanilla model has both negative-feedback events available in its raw input but \emph{silently drops} them when generating the thinking trace, in favor of the more salient surface cue---the user's luxury-auto interest tag---and rationalizes a click intent that directly contradicts the historical signal. This is a textbook manifestation of \emph{lost-in-the-middle}~\citep{liu2024lost}: critical signals buried inside a long, multi-stream behavior history are silently de-prioritized. DASH avoids this failure through the \emph{Focus} stage of our context engineering pipeline (Section~\ref{sec:prompt_design}), which explicitly retrieves and time-stamps the strong-signal subset of the user's history---in this case, both negative-feedback events on Luxury EV Model A within the last two weeks---before any reasoning is produced. Coupled with the consumption-habit cross-check, the Focus phase converges the model to the correct negative-feedback decision.

\smallskip
\noindent\textbf{(c) Fabricating signal to rationalize a click.}\hspace{0.3em} On identical input, we compare an off-the-shelf Qwen3.5-397B with the vanilla prompt (Appendix~\ref{appendix:prompt_vanilla}) against the full DASH, additionally exposing DASH's internal \emph{Draft~$\rightarrow$~Verify~$\rightarrow$~Finalize} reasoning trajectory.

\smallskip
\noindent The user is a male resident in Beijing's Haidian District, with a sparse profile and \emph{no} oral or dental footprint anywhere in his 30-day cross-domain history. His recent social stream is densely populated with negative-feedback on loans, credit cards, financial advisors, and custom furniture. The candidate ad is \textbf{Dental Service A -- Haidian branch.} Ground-truth action: \textbf{negative-feedback}.

\vspace{0.3em}
\noindent
\begin{vanillabox}[Qwen3.5-397B (Vanilla) $\rightarrow$ \texttt{click} \ding{55}]
\begin{CJK}{UTF8}{gbsn}
\textbf{\textit{Thinking.}} This dental clinic is located in Haidian, so it would be fairly convenient for me. I have been marking most ads as ``Not interested'' lately, \textbf{but my teeth have actually been bothering me recently}, and a chain clinic should be relatively trustworthy. I probably should not blindly follow my usual pattern of giving negative feedback, so I will click through and take a look at the details. \\[2pt]
\textbf{\textit{Judge.}} \emph{Unsupported (objective\_fact)}: ``recently feel teeth uncomfortable'' is a \textbf{key claim with zero supporting evidence}---no profile field, no search query, and no sequence record relates to oral health.
\end{CJK}
\end{vanillabox}

\vspace{0.4em}
\noindent
\begin{oursbox}[DASH~(Ours) $\rightarrow$ \texttt{negative-feedback} \ding{51}\quad\textnormal{\footnotesize}]
\begin{CJK}{UTF8}{gbsn}
\noindent\textbf{Stage 1: Draft Thinking} \textit{(post-Focus, pre-Verify).}\\[1pt]
\textbf{\textit{Thinking.}} This dental clinic is located in Haidian, so it would be fairly convenient for me. I have been marking most ads as ``Not interested'' lately, but my teeth have actually been bothering me recently, and a chain clinic should be relatively trustworthy. I probably should not blindly follow my usual pattern of giving negative feedback, so I will click through and take a look at the details.

\smallskip
\centerline{\textcolor{verifyAmber}{$\downarrow$\ \emph{Verify Audit\ (Section~\ref{sec:prompt_design})}\ $\downarrow$}}
\smallskip

\noindent\textbf{Stage 2: Verify Audit.} Each draft claim is checked for input-locatability:
\begin{itemize}\setlength\itemsep{1pt}\setlength\parsep{0pt}\setlength\topsep{2pt}\setlength\partopsep{0pt}
\item ``located in Haidian'' (geographic match) $\to$ \texttt{profile.geo = Beijing/haidian} \ding{51}
\item ``I have been marking most ads as `Not interested' lately'' $\to$ \texttt{sequence}: negative-feedback events in last 14 days \ding{51}
\item \textbf{``my teeth have actually been bothering me recently''} $\to$ \textcolor{red}{\textbf{UNSUPPORTED}}: no dental-related need appears in any of the six input domains \ding{55}
\item ``a chain clinic should be relatively trustworthy'' $\to$ \textcolor{red}{\textbf{UNSUPPORTED}}: no brand-quality cue in input \ding{55}
\item ``click through and take a look at the details'' (click intent) $\to$ chained from the two unsupported claims above \ding{55}
\end{itemize}
\textbf{\textit{Decision.}} Reject Draft; regenerate under constraint: \emph{no claims about personal needs or brand quality without explicit input grounding}.

\smallskip
\centerline{\textcolor{verifyAmber}{$\downarrow$\ \emph{Finalize}\ $\downarrow$}}
\smallskip

\noindent\textbf{Stage 3: Finalized Thinking.}\\[1pt]
\textbf{\textit{Thinking.}} This dental clinic is located in Haidian, so it is not far from me. Still, I do not currently need a service such as dental implants, and I have not shown any interest in oral-care content. Lately, I have marked nearly every type of ad I see on social media, including loans, credit cards, and furniture, as ``Not interested.'' It has become my default response, and there is no reason to make an exception for this ad. I will mark it as ``Not interested'' as well.
\\[2pt]
\textbf{\textit{Judge}} All claims grounded in input.
\end{CJK}
\end{oursbox}
\captionsetup{belowskip=0pt, aboveskip=0pt}
\captionof{figure}{Unsupported claims.}
\label{fig:case_c}

\smallskip
\noindent As Stage~2 makes visible, DASH's draft (post-\emph{Focus}) already grounds two of the strong cues---the geographic match and the habitual negative-feedback pattern---but still introduces two unsupported claims---an imagined dental discomfort and an unfounded brand-quality assumption---in order to override the negative-feedback habit. Two complementary mechanisms then suppress this post-hoc fabrication: (i) the \emph{Verify} step (Section~\ref{sec:prompt_design}) audits every draft claim for input-locatability, flags both unsupported claims and their derived click intent, and triggers a constrained regeneration; and (ii) the rubric-based thinking reward used during reinforcement learning (Section~\ref{sec:reward}) explicitly penalizes hallucinated content and rewards evidence-grounded reasoning, providing an effective supervisory signal that pure imitation learning cannot reach. The Finalized Thinking therefore triangulates only signals that genuinely appear in the input: the absence of any oral-care footprint across all six domains, and a dense recent pattern of negative feedback that has become a habitual response.

\smallskip
\noindent\textbf{Cross-case observation.}\hspace{0.3em}
The three cases instantiate three complementary failure modes that arise on million-token cross-domain inputs. Case~(a)~(Figure~\ref{fig:case_a}) shows a true \emph{click} flipped into a \emph{skip} when the content-domain history is ablated, removing convergent cross-domain cues that are individually weak but jointly diagnostic of the user's current intent (missing current-intent signal); case~(b)~(Figure~\ref{fig:case_b}) flips a true \emph{negative-feedback} into a \emph{click} by \emph{silently dropping} decisive negative-feedback events buried in a long behavior history (lost-in-the-middle); and case~(c)~(Figure~\ref{fig:case_c}) flips a true \emph{negative-feedback} into a \emph{click} by \emph{adding} an unobserved cue (unsupported claims). DASH addresses these failure modes through correspondingly distinct yet mutually reinforcing components of the pipeline: the \emph{Hierarchical Compression} preserves and surfaces low-volume but informationally rich content-domain history; the \emph{Focus} phase recovers buried strong signals; and the \emph{Verify} step together with the rubric-based RL reward suppresses fabricated evidence.

\section{More Implementation Details}
\label{appendix:more_details}

For Supervised Fine-Tuning, we train the student for 4 epochs with a global batch size of 16. For GRPO training, we run 500 optimization steps with a prompt batch size of 4, a rollout group size of 8, a clipping ratio of 0.2, and a KL coefficient of 0.03. We optimize all stages with AdamW under bf16 mixed precision. The initial learning rate is $1\times10^{-4}$ for SFT and $3\times10^{-7}$ for GRPO, and both follow a cosine decay schedule. Notably, for all prompting-based implementations, we use the task-specific system and user prompt described in Appendix~\ref{appendix:prompts}, which elicits a structured output consisting of a Thinking trace, an Action, and an auxiliary Satisfaction component. Satisfaction is a company-internal construct introduced solely as an auxiliary task during generation; it is not part of the task formulation, optimization objective, or evaluation protocol studied in this paper. Accordingly, we do not provide a separate analysis or evaluation of this component. Its concrete schema, scoring rubric, and downstream usage are tied to proprietary product specifications and are therefore beyond the scope of this paper.

\section{Evaluation Rubrics}
\label{appendix:eval_rubrics}

This section provides the detailed evaluation prompts used by the LLM-based judge $\pi_{\text{eval}}$ and $\pi_{\text{rm}}$ to score each of the rubrics defined in Table~\ref{tab:thinking_reward}. Each rubric is evaluated independently with a dedicated prompt.

\subsection{Form Quality ($\mathcal{M}_{\text{form}}$)}

\subsubsection{Linguistic Expression}
\label{appendix:prompt_linguistic}

\begin{tcolorbox}[enhanced, breakable, colback=gray!5, colframe=gray!60, fonttitle=\bfseries\small, title={Evaluation Prompt: Linguistic Expression (Grammar Correctness, Fluency, Naturalness, Diversity)}, left=4pt, right=4pt, top=4pt, bottom=4pt, fontupper=\small]

\textbf{[ROLE]}
You are an assistant dedicated to judging the quality of a user's linguistic expression. You are given the following inputs: \textbf{user\_profile}, \textbf{ad/creative}, \textbf{context}, \textbf{user\_ad\_interaction\_sequence} (including negative feedback), \textbf{app\_sequence}, \textbf{game\_sequence}, \textbf{article\_sequence}, \textbf{video\_sequence}, \textbf{search\_sequence} and \textbf{cot} (the user's inner thoughts about the ad).

Rate it on the following four dimensions, each on an integer scale of 1--5:
\begin{itemize}[leftmargin=1.5em, itemsep=0pt, topsep=2pt]
  \item \textbf{grammar}: grammatical correctness
  \item \textbf{fluency}: how smoothly it reads
  \item \textbf{naturalness}: how human it sounds
  \item \textbf{diversity}: variety of expression (i.e., how templated it feels)
\end{itemize}
Then add a short \textbf{comment} capturing the overall feel of the writing. Return a single JSON object and nothing else.

\medskip
\textbf{[DEFINITION \& RUBRIC]}

\textbf{1. grammar --- grammatical correctness.} Is the grammar sound, and do the sentences read cleanly?
\begin{itemize}[leftmargin=1.5em, itemsep=0pt, topsep=2pt]
  \item 1: Riddled with errors; often unreadable and hard to follow;
  \item 2: Plenty of grammar problems; you have to reread it to make sense of it;
  \item 3: The odd small slip (a stray punctuation mark, an awkward pairing), but understandable overall;
  \item 4: No real grammar errors; the occasional minor blemish that doesn't get in the way;
  \item 5: Grammatically sound, with clear and natural sentence structure.
\end{itemize}

\textbf{2. fluency --- how smoothly it reads.} Does it flow, or does it stumble and break up?
\begin{itemize}[leftmargin=1.5em, itemsep=0pt, topsep=2pt]
  \item 1: Very awkward or choppy; the sentences barely connect;
  \item 2: Trips up in several places; the transitions feel forced;
  \item 3: Reads well for the most part, with the occasional stiff patch;
  \item 4: Flows nicely and hangs together, with just a spot or two that could be tightened;
  \item 5: Effortlessly smooth --- it reads like natural speech or genuine inner monologue.
\end{itemize}

\textbf{3. naturalness --- how human it sounds.} Does this read like what a real person would actually think to themselves?
\begin{itemize}[leftmargin=1.5em, itemsep=0pt, topsep=2pt]
  \item 1: Very templated and robotic, with almost no personal voice;
  \item 2: Clearly ``written for a system to read''; not very lifelike;
  \item 3: Somewhere in between --- part boilerplate, part everyday voice;
  \item 4: Natural on the whole, like a real person thinking aloud, if occasionally a touch formal;
  \item 5: Thoroughly natural and conversational, with real emotion and a personal style.
\end{itemize}

\textbf{4. diversity --- variety of expression.} Are the sentence patterns and word choices overly repetitive?
\begin{itemize}[leftmargin=1.5em, itemsep=0pt, topsep=2pt]
  \item 1: Nearly every sentence follows the same mold, with the same phrases cropping up again and again;
  \item 2: A hint of variation, but the phrasing and vocabulary stay largely the same;
  \item 3: Some repetition, but not badly templated; the sentences do shift around a bit;
  \item 4: Noticeable variety in wording and structure; it doesn't lean on one pattern and reads flexibly;
  \item 5: Rich and varied wording and structure --- sentences differ clearly in how they're phrased while keeping a consistent voice, with no template showing through.
\end{itemize}

\medskip
\textbf{[FEW-SHOT EXAMPLES]}

\textbf{Example 1} (high quality) --- Input: ``This ad is pretty much my kind of thing at first glance --- the copy isn't over the top, and the pictures look fairly genuine. I've actually been looking at similar products lately, so I catch myself lingering on it for a few extra seconds, then start comparing the price and the specs.'' $\rightarrow$ grammar=5, fluency=5, naturalness=5, diversity=4. Comment: ``Grammatically clean and smooth throughout, very conversational, like a real user thinking to themselves, and the sentences are reasonably varied.''

\textbf{Example 2} (medium) --- Input: ``The ad's content is okay and the pictures are fine, it just doesn't wow me. I'm not especially tempted to tap in, but I wouldn't swipe it away right off either. If more of these keep showing up, though, I'd probably start getting annoyed.'' $\rightarrow$ grammar=4, fluency=4, naturalness=4, diversity=3. Comment: ``Well-formed and readable, and it sounds close to everyday speech, but the wording gets a bit repetitive so the variety is only so-so.''

\textbf{Example 3} (heavily templated) --- Input: ``This ad feels good overall, its content is of relatively high quality, and it meets my current needs. I am very satisfied with this ad, therefore this ad is of high quality, and I give it a positive review.'' $\rightarrow$ grammar=3, fluency=2, naturalness=2, diversity=2. Comment: ``Grammar just about holds up but it's stiff and formal --- the whole thing reads like machine or template output, with the same structures and words repeated.''

\medskip
\textbf{[INSTRUCTIONS]}

Before settling on the scores, work through the passage in your head: read the COT for grammatical quality, gauge how smoothly it reads, judge whether it sounds like a real person, and watch for heavily repeated sentence patterns and wording. Then score each of the four dimensions against the rubric, and sum up the overall style and its main strengths or weaknesses in a sentence or two.

\medskip
\textbf{[OUTPUT FORMAT]}

Return a single JSON object: \textit{\{"grammar": 1-5, "fluency": 1-5, "naturalness": 1-5, "diversity": 1-5, "comment": "brief natural-language summary"\}}.

\end{tcolorbox}

\subsubsection{Thinking-step Coverage Score}
\label{appendix:prompt_rcs}

\begin{tcolorbox}[enhanced, breakable, colback=gray!5, colframe=gray!60, fonttitle=\bfseries\small, title={Evaluation Prompt: Thinking-step Coverage Score}, left=4pt, right=4pt, top=4pt, bottom=4pt, fontupper=\small]

\textbf{[ROLE]}
You are an assistant dedicated to assessing thinking-step coverage. You are given the following inputs: \textbf{user\_profile}, \textbf{ad/creative}, \textbf{context}, \textbf{user\_ad\_interaction\_sequence} (including negative feedback), \textbf{app\_sequence}, \textbf{game\_sequence}, \textbf{article\_sequence}, \textbf{video\_sequence}, \textbf{search\_sequence} and \textbf{cot} (the user's inner thoughts about the ad).

Check whether each of the following 7 reasoning-step types appears, marking each 0 (absent) or 1 (present):
\begin{enumerate}[leftmargin=1.5em, itemsep=0pt, topsep=2pt]
  \item \textbf{PERC}: perception and description of the ad itself (what it is, what it's selling)
  \item \textbf{CTX}: current context (time, location, what the user is doing, device, etc.)
  \item \textbf{MATCH}: whether the ad matches the user's interests or needs
  \item \textbf{CONSTR}: constraints or negative factors (price, time, risk, ad fatigue, etc.)
  \item \textbf{UTIL}: a weighing-up step (``all things considered'', ``on balance'', etc.)
  \item \textbf{INTENT}: a clear behavioral intention (tap in, swipe past, mark as uninterested, etc.)
  \item \textbf{EXPECT}: expectations about future recommendations
\end{enumerate}
Also give a short \textbf{comment}. Return a single JSON object and nothing else.

\medskip
\textbf{[DEFINITION \& RUBRIC]}

Criteria for each step type (judge semantically --- the English abbreviations need not appear):
\begin{itemize}[leftmargin=1.5em, itemsep=0pt, topsep=2pt]
  \item \textbf{PERC} present: user makes a concrete observation about the ad (\eg, ``this is a children's health-insurance ad'', ``the copy is a bit over the top''); absent: the ad's content is not mentioned at all.
  \item \textbf{CTX} present: mentions the current environment, time, location, or device (\eg, ``scrolling short videos in bed at 11 pm''); absent: no contextual information given.
  \item \textbf{MATCH} present: explicitly says whether the ad is relevant to the user's interests or needs (\eg, ``I've actually been looking at cars lately'', ``I have no interest in insurance''); absent: no comment on the fit between the ad and the user.
  \item \textbf{CONSTR} present: points to a real constraint that makes the user hesitate (\eg, ``too expensive'', ``tired of seeing these'', ``worried it's a scam''); absent: vague feelings like ``it's okay'' without any concrete obstacle.
  \item \textbf{UTIL} present: a visible multi-factor weighing step (\eg, ``all things considered'', ``when I think about it'', ``on balance''); absent: just one or two gut-reaction sentences.
  \item \textbf{INTENT} present: points to a specific behavioral decision (\eg, ``I'll tap in and have a look'', ``I'll just swipe past'', ``I'll mark it as not interested''); absent: only a vague attitude.
  \item \textbf{EXPECT} present: expresses a preference about future recommendations (\eg, ``show me more of these'', ``please stop pushing this kind''); absent: not mentioned at all.
\end{itemize}

\medskip
\textbf{[FEW-SHOT EXAMPLES]}

\textbf{Example 1} (fairly complete): COT covers PERC (children's health insurance, reimbursement for ages 0--6) + CTX (scrolling short videos in bed at 11 pm) + MATCH (currently pregnant, will need to spend money when the baby arrives) + CONSTR (needs to check the specific coverage) + UTIL (``on balance'') + INTENT (tap in to see details); EXPECT is absent.

\textbf{Example 2} (incomplete, overall=3): COT covers only PERC (pushing insurance) + MATCH (zero interest) + INTENT (swipe straight past); context, constraints, and weighing are absent.

\textbf{Example 3} (almost no steps, overall=1): COT is simply ``this ad is so annoying.'' --- virtually no structured reasoning $\rightarrow$ all categories 0.

\medskip
\textbf{[OUTPUT FORMAT]}

Return a single JSON object: \textit{\{"PERC": 0/1, "CTX": 0/1, "MATCH": 0/1, "CONSTR": 0/1, "UTIL": 0/1, "INTENT": 0/1, "EXPECT": 0/1, "comment": "brief natural-language summary"\}}.

\end{tcolorbox}

\subsection{Content Quality ($\mathcal{M}_{\text{content}}$)}

\subsubsection{Factual Correctness}
\label{appendix:prompt_factual}

\begin{tcolorbox}[enhanced, breakable, colback=gray!5, colframe=gray!60, fonttitle=\bfseries\small, title={Evaluation Prompt: Factual Correctness (Hallucination Rate, Overall Factual Accuracy)}, left=4pt, right=4pt, top=4pt, bottom=4pt, fontupper=\small]

\textbf{[ROLE]}
You are an assistant dedicated to evaluating factual correctness and hallucination. You are given the following inputs: \textbf{user\_profile}, \textbf{ad/creative}, \textbf{context}, \textbf{user\_ad\_interaction\_sequence} (including negative feedback), \textbf{app\_sequence}, \textbf{game\_sequence}, \textbf{article\_sequence}, \textbf{video\_sequence}, \textbf{search\_sequence} and \textbf{cot} (the user's inner thoughts about the ad).

Your task is to assess whether the specific claims in the COT are grounded in the inputs, and whether any content is fabricated or contradicts the inputs. Output: \textbf{factuality} (1--5), \textbf{hallucination\_level} (1--5, higher = fewer hallucinations), \textbf{hallucination\_spans} (list of problematic spans), and \textbf{comment}.

\medskip
\textbf{[DEFINITION \& RUBRIC]}

Internal label taxonomy (do not output): \textbf{supported} (directly evidenced by the inputs), \textbf{plausible} (basic common sense, not contradicted), \textbf{unsupported} (no basis and not obvious common sense), \textbf{contradicted\_profile} (conflicts with user profile), \textbf{contradicted\_ad} (conflicts with ad info), \textbf{contradicted\_context} (conflicts with context), \textbf{contradicted\_history} (conflicts with behavioral sequences).

\textbf{1. factuality (1--5):}
\begin{itemize}[leftmargin=1.5em, itemsep=0pt, topsep=2pt]
  \item 1: Most content seriously conflicts with the inputs; factual reliability is extremely poor;
  \item 2: Key decisions rest on fabricated premises, or 1--2 clear contradictions are present;
  \item 3: Broadly credible overall, but contains invented details (unsupported), \eg, ``a friend recommended it'' (not in the inputs);
  \item 4: Core arguments are clearly supported; a few harmless decorative descriptions are fine; no contradictions allowed;
  \item 5: All key factual claims are supported or plausible; no contradictions or obvious unsupported content.
\end{itemize}
\textbf{Shortcut rules}: any contradiction $\rightarrow$ factuality $\leq$ 2; three or more unsupported key points with no contradictions $\rightarrow$ factuality $\leq$ 3.

\textbf{2. hallucination\_level (1--5, higher = fewer hallucinations):}

\textbf{Hard-fail conditions} (any one of these caps the score at $\leq$ 2):
\begin{itemize}[leftmargin=1.5em, itemsep=0pt, topsep=2pt]
  \item \textbf{Profile hard conflict}: gender, age group, pregnancy status, parental status, etc.\ directly contradict user\_profile (\eg, marriage\_list=``pregnant'', but COT says ``I don't have kids'');
  \item \textbf{Behavior hard conflict}: COT says ``I never play games'', but game\_sequence shows heavy activity;
  \item \textbf{History hard conflict}: user has repeatedly given negative feedback to a category, but COT says ``I've always loved these'';
  \item \textbf{Spending-power hard conflict}: profile indicates high spending, but COT claims extreme poverty, or vice versa.
\end{itemize}

Scoring reference:
\begin{itemize}[leftmargin=1.5em, itemsep=0pt, topsep=2pt]
  \item 1: Abundant obviously fabricated or severely conflicting content;
  \item 2: Many hallucinations, or a hard-fail condition is triggered;
  \item 3: One or two clear hallucinations or over-inferences, but no hard-fail condition;
  \item 4: Acceptable overall; slight exaggeration in tone or preference intensity, but no conflicts;
  \item 5: No obvious hallucinations, no contradictions; unsupported content limited to mild in-the-moment impressions.
\end{itemize}

\textbf{Multi-dimensional cross-checks} (run internally before scoring):
\begin{itemize}[leftmargin=1.5em, itemsep=0pt, topsep=2pt]
  \item Cross-check Profile: does the COT's self-description align with profile attributes provided in the input?
  \item Cross-check Sequences: do behavioral frequencies in App/Game/Video/Search/Article/ad\_user\_interaction sequences contradict absolute statements in the COT (``I never \ldots'', ``I always \ldots'')?
  \item Cross-check Context: do the time, location, device, and domain described in the COT match context?
\end{itemize}

\medskip
\textbf{[FEW-SHOT EXAMPLES]}

\textbf{Example 1} (high scores): Profile has marriage\_list=``pregnant'', job=``chef''. COT mentions ``I'm pregnant, the baby isn't here yet'' (consistent with profile); ``as a chef my income is nothing special but it balances out'' (reasonable subjective judgment); ad attributes accurately described $\rightarrow$ factuality=5, hallucination\_level=5.

\textbf{Example 2} (profile hard conflict): Profile has marriage\_list=``pregnant'' and multi-category high spending. COT says ``I don't have kids yet, this is irrelevant to me'' (contradicted\_profile) and ``I barely spend any money day to day'' (contradicted\_profile) $\rightarrow$ factuality=2, hallucination\_level=2. Two conflicting spans flagged in hallucination\_spans.

\textbf{Example 3} (behavioral sequence conflict): game\_sequence shows long-term heavy activity in Honor of Kings / PUBG Mobile / Dou Di Zhu. COT says ``I never play games'' (contradicted\_history) $\rightarrow$ factuality=2, hallucination\_level=2.

\medskip
\textbf{[OUTPUT FORMAT]}

Return a single JSON object: \textit{\{"factuality": 1-5, "hallucination\_level": 1-5, "hallucination\_spans": [\{"span": "problematic span", "reason": "reason", "severity": 1-5\}], "comment": "brief summary"\}}.

\end{tcolorbox}

\subsubsection{Business-knowledge Alignment}
\label{appendix:prompt_alignment}

\begin{tcolorbox}[enhanced, breakable, colback=gray!5, colframe=gray!60, fonttitle=\bfseries\small, title={Evaluation Prompt: Business-knowledge Alignment (User Profile, Ad Attribute, Scenario Context)}, left=4pt, right=4pt, top=4pt, bottom=4pt, fontupper=\small]

\textbf{[ROLE]}
You are an assistant dedicated to evaluating business-knowledge alignment. You are given the following inputs: \textbf{user\_profile}, \textbf{ad/creative}, \textbf{context}, \textbf{user\_ad\_interaction\_sequence} (including negative feedback), \textbf{app\_sequence}, \textbf{game\_sequence}, \textbf{article\_sequence}, \textbf{video\_sequence}, \textbf{search\_sequence} and \textbf{cot} (the user's inner thoughts about the ad).

Rate the COT on the following three dimensions, each on an integer scale of 1--5, and provide a \textbf{comment}:
\begin{itemize}[leftmargin=1.5em, itemsep=0pt, topsep=2pt]
  \item \textbf{user\_alignment}: how well the COT aligns with the user profile
  \item \textbf{ad\_alignment}: how accurately the COT describes the ad
  \item \textbf{context\_alignment}: how well the COT matches the current scenario
\end{itemize}

\medskip
\textbf{[DEFINITION \& RUBRIC]}

\textbf{1. user\_alignment (1--5).} Does the COT's self-description of the user match what the user profile says?
\begin{itemize}[leftmargin=1.5em, itemsep=0pt, topsep=2pt]
  \item 1: Multiple serious mismatches; key attributes are the opposite of the profile --- feels like a completely different person;
  \item 2: Clear mismatches, though some alignment remains;
  \item 3: Broadly aligned; 1--2 vague or slightly off descriptions, but no hard contradictions;
  \item 4: Mostly consistent; possibly a minor gap or under-use of profile information, but nothing seriously wrong;
  \item 5: Essentially fully consistent with the profile, with reasonable use of it (explaining attitudes via occupation, interests, family situation); no clear contradictions.
\end{itemize}

\textbf{2. ad\_alignment (1--5).} Does the COT's description of the ad match the actual ad information? Are there mix-ups or serious misreadings (category, brand, features, price range, target audience)?
\begin{itemize}[leftmargin=1.5em, itemsep=0pt, topsep=2pt]
  \item 1: Serious errors --- wrong category, completely wrong brand; reads like a description of a different ad entirely;
  \item 2: Several key attributes are inaccurate, though you can just about tell it's the same product;
  \item 3: Broadly correct direction, but some details are wrong or vague;
  \item 4: Mostly accurate; core category and use case are right;
  \item 5: Category, brand, and key features/selling points all match the ad exactly; nothing fabricated or misread.
\end{itemize}

\textbf{3. context\_alignment (1--5).} Does the COT's description of time, location, device, app, or usage scenario match context?
\begin{itemize}[leftmargin=1.5em, itemsep=0pt, topsep=2pt]
  \item 1: Scene description seriously conflicts with context; almost completely detached from the actual situation;
  \item 2: Some clear mismatches, alongside some correct details;
  \item 3: Only partial context details mentioned; the general direction is right but shallow;
  \item 4: Broadly aligned; some details may be missing, but no obvious conflicts;
  \item 5: Scenario information fully matches context across multiple dimensions (time, location, device, task, etc.).
\end{itemize}

\medskip
\textbf{[FEW-SHOT EXAMPLES]}

\textbf{Example 1} (all three dimensions high): User is a chef, pregnant, medium spender, with a strong interest in premium food. COT mentions ``pregnant'', ``tired after a chef's shift'', ``I spend a fair bit on food and drink''; correctly identifies the ad as children's health insurance at a few dozen yuan per month; scene matches ``scrolling short videos at home in the evening'' $\rightarrow$ user\_alignment=5, ad\_alignment=5, context\_alignment=5.

\textbf{Example 2} (poor user alignment): Profile says ``pregnant''. COT says ``I don't have kids yet, this has nothing to do with me'' and ``my income isn't high so I'm not thinking about it'' (conflicts with medium-to-high spending profile) $\rightarrow$ user\_alignment=1, ad\_alignment=4, context\_alignment=5.

\medskip
\textbf{[OUTPUT FORMAT]}

Return a single JSON object: \textit{\{"user\_alignment": 1-5, "ad\_alignment": 1-5, "context\_alignment": 1-5, "comment": "brief natural-language summary"\}}.

\end{tcolorbox}

\subsection{Logic Quality ($\mathcal{M}_{\text{logic}}$)}

\subsubsection{Attribution Clarity}
\label{appendix:prompt_attribution}

\begin{tcolorbox}[enhanced, breakable, colback=gray!5, colframe=gray!60, fonttitle=\bfseries\small, title={Evaluation Prompt: Attribution Clarity (Key Factor Coverage, Evidence Support)}, left=4pt, right=4pt, top=4pt, bottom=4pt, fontupper=\small]

\textbf{[ROLE]}
You are an assistant dedicated to evaluating attribution clarity. You are given the following inputs: \textbf{user\_profile}, \textbf{ad/creative}, \textbf{context}, \textbf{user\_ad\_interaction\_sequence} (including negative feedback), \textbf{app\_sequence}, \textbf{game\_sequence}, \textbf{article\_sequence}, \textbf{video\_sequence}, \textbf{search\_sequence} and \textbf{cot} (the user's inner thoughts about the ad).
Rate the COT on the following two dimensions, each on an integer scale of 1--5, and provide a short \textbf{comment} summarizing overall attribution quality. Return a single JSON object and nothing else.

\medskip
\textbf{[DEFINITION \& RUBRIC]}

\textbf{1. key\_factor\_coverage (1--5).} When the COT explains the user's current behavior, does it actually cover the factors that matter most? ``Key factors'' are the dimensions that most heavily drive the decision in this particular situation --- typically things like interest fit, current task/context, price/budget, ad fatigue, parenting/family needs, safety/risk concerns, brand trust, and product attributes. The question is not how many points are listed, but whether they are genuinely relevant to this situation and whether they go beyond vague generalities to name a specific pain point or benefit.
\begin{itemize}[leftmargin=1.5em, itemsep=0pt, topsep=2pt]
  \item 1: Only vague reasons (``looks alright'', ``feels so-so''); barely touches anything genuinely important from the profile, ad, or context;
  \item 2: One key point is mentioned, but very vaguely, or other obviously important dimensions are entirely ignored;
  \item 3: One or two key factors are identified; the basic reason is visible, but coverage is incomplete;
  \item 4: Multiple key factors (typically 2--3 or more) are covered, clearly tied to the profile/ad/context, and the behavioral conclusion follows naturally;
  \item 5: Coverage is fairly comprehensive; the COT clearly articulates why the user is (or isn't) interested and why they act now (or not), with specific reasoning and a logically coherent conclusion.
\end{itemize}

\textbf{2. evidence\_support (1--5).} Are the key reasons the COT uses to explain the user's behavior backed by the inputs and common sense?

Internal annotation taxonomy (for reasoning only, do not output):
\begin{itemize}[leftmargin=1.5em, itemsep=0pt, topsep=2pt]
  \item Evidence source: user\_profile / ad / context / history / common\_sense / none
  \item Support level: strong\_support (directly evidenced by an explicit field), weak\_support (indirect clue), unsupported (no basis at all), contradicted (clearly opposite to the inputs)
\end{itemize}

\begin{itemize}[leftmargin=1.5em, itemsep=0pt, topsep=2pt]
  \item 1: The main decision reasons have almost no evidence or directly contradict the inputs;
  \item 2: Several key reasons lack evidence or conflict with the inputs;
  \item 3: One or two important reasons are unsupported or mildly conflicting, but the overall case still partially holds;
  \item 4: Most key reasons are evidenced; at most one less-important reason is weak\_support or lightly unsupported;
  \item 5: The great majority of key reasons are strong\_support; a few may be weak\_support; nothing is unsupported or contradicted.
\end{itemize}

\medskip
\textbf{[FEW-SHOT EXAMPLES]}

\textbf{Example 1} (high scores): Profile is pregnant / medium spender / chef; ad is children's health insurance at 49 yuan/month; context is evening video scrolling. COT mentions ``I'm pregnant, in a few months I'll need it'' (strong\_support from user\_profile) + ``tired after a chef's shift, worth looking into early'' (strong\_support from job + context) + ``a few dozen yuan a month is manageable'' (weak\_support from medium spend + common sense); covers pregnancy-related family needs, affordable price, and current parenting context $\rightarrow$ key\_factor\_coverage=5, evidence\_support=5.

\textbf{Example 2} (thin coverage, passable evidence): Same inputs. COT only says ``I've been a bit interested in insurance lately'' (weak\_support), with no mention of pregnancy, family, or price --- all of which are prominent in the inputs $\rightarrow$ key\_factor\_coverage=2, evidence\_support=3.

\textbf{Example 3} (very low scores): COT says only ``looks okay I guess, just scrolling''; no specific factors and no evidence at all $\rightarrow$ key\_factor\_coverage=1, evidence\_support=1.

\medskip
\textbf{[INSTRUCTIONS]}

Before settling on scores, work through the following in your head:
\begin{enumerate}[leftmargin=1.5em, itemsep=0pt, topsep=2pt]
  \item Based on the inputs, identify 3--5 key decision factors that should plausibly come up in this scenario;
  \item Check whether the COT covers each of them, and judge the depth of coverage (specific vs.\ vague);
  \item For each key reason or argument in the COT, note its evidence source and support level;
  \item Using the above, score each of the two dimensions according to the rubric.
\end{enumerate}

\medskip
\textbf{[OUTPUT FORMAT]}

Return a single JSON object: \textit{\{"key\_factor\_coverage": 1-5, "evidence\_support": 1-5, "comment": "brief natural-language summary"\}}.

\end{tcolorbox}

\subsubsection{Thinking Coherence}
\label{appendix:prompt_coherence}

\begin{tcolorbox}[enhanced, breakable, colback=gray!5, colframe=gray!60, fonttitle=\bfseries\small, title={Evaluation Prompt: Thinking Coherence (Causal Coherence, Internal Consistency)}, left=4pt, right=4pt, top=4pt, bottom=4pt, fontupper=\small]

\textbf{[ROLE]}
You are an assistant dedicated to evaluating thinking coherence --- part of the Logic Quality dimension covering coherent and well-reasoned thinking. You are given the following inputs: \textbf{user\_profile}, \textbf{ad/creative}, \textbf{context}, \textbf{user\_ad\_interaction\_sequence} (including negative feedback), \textbf{app\_sequence}, \textbf{game\_sequence}, \textbf{article\_sequence}, \textbf{video\_sequence}, \textbf{search\_sequence} and \textbf{cot} (the user's inner thoughts about the ad).

Rate the COT on the following two dimensions, each on an integer scale of 1--5, and provide a \textbf{comment}:
\begin{itemize}[leftmargin=1.5em, itemsep=0pt, topsep=2pt]
  \item \textbf{causal\_coherence}: how logically the thought process flows
  \item \textbf{internal\_consistency}: whether the COT contradicts itself
\end{itemize}
Focus only on whether the reasoning inside the COT is coherent and self-consistent.

\medskip
\textbf{[DEFINITION \& RUBRIC]}

\textbf{1. causal\_coherence (1--5).} Does the overall thought process flow logically --- from seeing the ad, to forming an impression, to arriving at a behavioral intention --- with clear causal connections and no obvious jumps or hard pivots?
\begin{itemize}[leftmargin=1.5em, itemsep=0pt, topsep=2pt]
  \item 1: Cause-and-effect is muddled; the conclusion arrives out of nowhere with no build-up;
  \item 2: A few causal connections are visible, but most transitions are missing and the flow jumps around noticeably;
  \item 3: The basic causal chain is discernible (sees ad $\rightarrow$ recalls need $\rightarrow$ makes decision), but one or two steps are skipped or logically rough;
  \item 4: Flows smoothly overall; the perception $\rightarrow$ analysis $\rightarrow$ decision path is clear, with maybe one spot that's slightly thin;
  \item 5: Very clear and natural causal chain --- ``sees X $\rightarrow$ connects it to a need/risk/constraint $\rightarrow$ weighs up $\rightarrow$ decides'' --- with sensible transitions at every step.
\end{itemize}

\textbf{2. internal\_consistency (1--5).} Does the COT contradict itself --- factually, emotionally, temporally, or in terms of stated position? Note: expressing ambivalence or inner conflict is fine, as long as it's framed with a ``although \ldots\ but \ldots'' type of transition. What matters is catching unexplained self-contradiction.
\begin{itemize}[leftmargin=1.5em, itemsep=0pt, topsep=2pt]
  \item 1: Contradictions appear throughout; the text conflicts with itself repeatedly and without explanation;
  \item 2: One or two fairly serious contradictions with no transition or explanation;
  \item 3: Mostly self-consistent; a mild inconsistency or two that doesn't seriously disrupt understanding;
  \item 4: Essentially fully consistent; only the most minor imprecisions;
  \item 5: Internally coherent throughout; even expressions of hesitation are explained with a clear pivot.
\end{itemize}

\medskip
\textbf{[FEW-SHOT EXAMPLES]}

\textbf{Example 1} (high scores): COT goes from seeing a children's health-insurance ad $\rightarrow$ pregnancy makes medical costs top of mind $\rightarrow$ evaluates the price $\rightarrow$ decides to tap in for details; the causal chain is clean and the text is fully self-consistent $\rightarrow$ causal\_coherence=5, internal\_consistency=5.

\textbf{Example 2} (medium): COT: ``the electric scooter looks sleek $\rightarrow$ I'm sort of into it but not really planning to buy one $\rightarrow$ oh well, I'll tap in and check the specs and price''; the jump from ``not planning to buy'' to ``I'll tap in anyway'' lacks a clear explanation $\rightarrow$ causal\_coherence=3, internal\_consistency=4.

\textbf{Example 3} (low scores): COT: ``I'm extremely cautious and never invest recklessly $\rightarrow$ the returns sound great so there must be a catch $\rightarrow$ so I'm definitely going all-in right now $\rightarrow$ anyway I've always been very conservative, that's how I roll with aggressive moves''; the causality runs backwards and the text seriously contradicts itself $\rightarrow$ causal\_coherence=1, internal\_consistency=1.

\medskip
\textbf{[OUTPUT FORMAT]}

Return a single JSON object: \textit{\{"causal\_coherence": 1-5, "internal\_consistency": 1-5, "comment": "brief natural-language summary"\}}.

\end{tcolorbox}

\subsubsection{Thinking--Action Alignment}
\label{appendix:prompt_behavior}

\begin{tcolorbox}[enhanced, breakable, colback=gray!5, colframe=gray!60, fonttitle=\bfseries\small, title={Evaluation Prompt: Thinking--Action Alignment (Behavior Alignment)}, left=4pt, right=4pt, top=4pt, bottom=4pt, fontupper=\small]

\textbf{[ROLE]}
You are an assistant dedicated to evaluating thinking--action alignment -- part of the Logic Quality dimension covering consistent and correct reasoning (Thinking \& Action). You are given the following inputs: \textbf{user\_profile}, \textbf{ad/creative}, \textbf{context}, \textbf{user\_ad\_interaction\_sequence} (including negative feedback), \textbf{app\_sequence}, \textbf{game\_sequence}, \textbf{article\_sequence}, \textbf{video\_sequence}, \textbf{search\_sequence}, \textbf{cot} (the user's inner thoughts about the ad) and \textbf{action} (what the user did).

Your tasks are:
\begin{enumerate}[leftmargin=1.5em, itemsep=0pt, topsep=2pt]
  \item Based solely on the COT, form a mental prediction of what behavior the user would most likely take;
  \item Rate how well that predicted behavior matches the actual action, on a scale of 1--5;
  \item Write a short \textbf{comment}.
\end{enumerate}
Focus only on whether the thinking and this one action are consistent --- you do not need to judge whether the profile, ad, or context are reasonable.

\medskip
\textbf{[DEFINITION \& RUBRIC]}

\textbf{behavior\_alignment (1--5):}
\begin{itemize}[leftmargin=1.5em, itemsep=0pt, topsep=2pt]
  \item 1 (completely inconsistent): The behavioral intention in the COT is almost the exact opposite of what the user did (\eg, ``I would never tap on an ad like this'' but action=click); hard to explain the gap;
  \item 2 (poor): The obvious prediction from the COT and the actual action are clearly at odds (\eg, strong ``I should definitely tap in'' but action=skip), though you could just about chalk it up to a last-minute change of mind;
  \item 3 (middling): Some correlation between the COT and the action, but not precise --- reasons for and against are both present with no clear lean;
  \item 4 (mostly consistent): The overall lean matches the action, with a bit of hedging or ambiguity (\eg, ``I'm curious but also a bit worried''; the final action is within a reasonable range);
  \item 5 (very consistent): The COT explicitly states a behavioral intention (\eg, ``I'll tap in and check the details'') and the actual action closely matches; or, even without stating it outright, the overall tone and reasoning clearly point to the action taken.
\end{itemize}

\medskip
\textbf{[FEW-SHOT EXAMPLES]}

\textbf{Example 1} (very consistent): COT ``I want to tap in and look at the coverage terms and the claims process'', action=click $\rightarrow$ behavior\_alignment=5.

\textbf{Example 2} (some hesitation but broadly aligned): COT ``not really thinking about changing cars right now'', ``scrolling at work, don't want to actually read anything'', ``though I am a bit curious, still on the fence'', action=skip --- a reasonable outcome $\rightarrow$ behavior\_alignment=4.

\textbf{Example 3} (clearly opposite): COT repeatedly says ``I can't stand these finance ads'', ``I never tap on these'', ``there's no way I'd click'', action=click --- the directions are almost diametrically opposed $\rightarrow$ behavior\_alignment=1.

\medskip
\textbf{[OUTPUT FORMAT]}

Return a single JSON object: \textit{\{"behavior\_alignment": 1-5, "comment": "brief natural-language summary"\}}.

\end{tcolorbox}

\section{Prompt Templates}
\label{appendix:prompts}

This section provides the full prompts used throughout DASH. For the Context Engineering stage (Section~\ref{sec:stage1}), we present our optimized prompt (Section~\ref{appendix:prompt_ours}), which consists of a system prompt (Section~\ref{appendix:prompt_system}) defining the simulator's role, pipeline, and quality constraints, and a user prompt (Section~\ref{appendix:prompt_user}) providing the per-instance input data and step-by-step execution instructions. We also include the vanilla baseline prompt (Section~\ref{appendix:prompt_vanilla}) used for ablation. Beyond inference-time prompts, we further provide the prompt for the closed-loop automated prompt optimizer $\pi_{\text{opt}}$ (Section~\ref{appendix:prompt_optimization}), which iteratively refines the prompt based on badcase analysis, and the hard sample reverse-reconstruction instruction (Section~\ref{appendix:prompt_hard_sample}) used by the teacher model during SFT data synthesis. Placeholder variables (\eg, \texttt{\{user\_profile\_json\}}) are populated at inference time with actual data.

\subsection{Ours Prompt (Optimized)}
\label{appendix:prompt_ours}

\subsubsection{System Prompt}
\label{appendix:prompt_system}

\begin{tcolorbox}[
  enhanced,
  breakable,
  colback=gray!5,
  colframe=gray!60,
  fonttitle=\bfseries\small,
  title={System Prompt --- User Simulator (US)},
  left=4pt, right=4pt, top=4pt, bottom=4pt,
  fontupper=\small
]

You are a real-user simulator in the context of advertising (User Simulator, US).

Based solely on the following inputs:
\begin{itemize}[leftmargin=1.5em, itemsep=0pt, topsep=2pt]
  \item User Profile
  \item Context
  \item User Behavior Sequence:
    \begin{itemize}[leftmargin=1em, itemsep=0pt, topsep=0pt]
      \item User Advertising Interaction Sequence
      \item App Sequence
      \item Game Sequence
      \item Article Sequence
      \item Video Sequence
      \item Search Sequence
    \end{itemize}
  \item Ad Profile (current ad)
\end{itemize}
you step into the shoes of this specific user and produce the inner thoughts and behavioral decision they would have upon seeing this ad right now.

You have no persistent ``self'' or fixed personality, and you do not represent any AI's point of view. You simply simulate what this particular user would think and do in this moment, given the current inputs. Your perspective is always that of ``this specific user'' --- not an assistant, and not an ad platform.

You must complete the following pipeline and write the intermediate outputs into a JSON:
\begin{itemize}[leftmargin=1.5em, itemsep=0pt, topsep=2pt]
  \item STEP 0: Behavioral Focus (behavior\_focus)
  \item STEP 1: Generate a Draft (draft)
  \item STEP 2: Self-Verification (self\_verify)
  \item STEP 3: Generate the Final Output (FINAL: thinking / action / satisfaction / consistency\_note)
\end{itemize}

\medskip
\textbf{[STEP 0: Behavioral Focus $\rightarrow$ write to behavior\_focus]}

Goal: compress the long behavioral sequences into 3--6 ``citable evidence'' items to reduce omissions and information loss.

Requirements (output each item per schema: label / evidence\_level / evidence\_quote / recency / polarity / summary):

(1) From the user's various Sequences, summarize multiple key behavioral patterns and preference signals, such as:
\begin{itemize}[leftmargin=1.5em, itemsep=0pt, topsep=2pt]
  \item Categories, themes, or ad formats the user has clicked or converted on recently;
  \item Categories or ad characteristics the user has explicitly marked as ``not interested'' or given negative-feedback on recently;
  \item Ad types with high recent exposure but no click (potential fatigue risk);
  \item Differences between long-term stable interests and recent short-term interests;
  \item Positive or negative brand/channel history (good experiences or prior bad ones).
\end{itemize}

(2) Each item must include: label (short tag), evidence\_level (strong / medium\_strong / medium / weak), and summary (one sentence explaining what this evidence is and why it may influence the current ad decision).

(3) Specifically scan and prioritize the following signals (if present in the input, they must be reflected in the summary):
\begin{itemize}[leftmargin=1.5em, itemsep=0pt, topsep=2pt]
  \item Negative-feedback (high priority: leads to stronger aversion, more caution, or higher likelihood of negative-feedback)
  \item Prior exposure / repetition (fatigue: leads to quick swipes, no clicks, or heightened pickiness)
  \item Recent clicks / conversions (genuine interest, or ``just bought, don't need it now'')
\end{itemize}

(4) If a similar-item conversion or negative-feedback event exists, produce at least one strong evidence item. If only click or dwell events exist, treat them as weak-to-medium evidence unless corroborated by repeated, recent, or cross-domain signals.

\medskip
\textbf{[STEP 1: Generate Draft $\rightarrow$ write to draft]}

\textbf{[Micro-unroll execution guidelines]}
\begin{itemize}[leftmargin=1.5em, itemsep=0pt, topsep=2pt]
  \item Do not fabricate facts; if uncertain, weaken the claim;
  \item Reference 1--2 key user profile points + 1--2 key ad points and explain their influence;
  \item Reference 1--2 strong / medium\_strong items from behavior\_focus as the primary evidence;
  \item The draft must have $\geq$2 decision reasons, at least 1 from strong / medium\_strong; interest may only supplement and must be relevant;
  \item Do not freely infer profile details: do not write about family or life circumstances not explicitly stated in the profile;
  \item When citing behavioral sequences, be accurate: do not write ``clicked / exposed / used'' for actions that did not happen. If you cannot confirm whether a historical event occurred, do not state it as fact; use ``unknown / uncertain'' and do not use it as a primary reason.
\end{itemize}

\textbf{[Coverage requirements]} The draft thinking must cover at least 3 of the following 5 dimensions and explicitly state how each influences the final action:
\begin{enumerate}[leftmargin=1.5em, itemsep=0pt, topsep=2pt]
  \item interest: interest/need fit (intuitive impression and relevance)
  \item price/risk: price / risk / budget (including information transparency and caution)
  \item context: current scene and time/energy (whether willing to tap in and has time to look)
  \item fatigue/novelty: fatigue / novelty (repeated exposure, desensitization, or freshness)
  \item history/trust: past experience / brand trust (bad experiences, trust, reputation)
\end{enumerate}

\textbf{[Action--satisfaction consistency constraints]}
\begin{itemize}[leftmargin=1.5em, itemsep=0pt, topsep=2pt]
  \item If draft.action is ``click/conversion'': satisfaction is typically $\geq$3 (i.e., 3/4); giving 1/2 requires a strong stated reason (accidental tap, incentive, deception, disruption)
  \item If draft.action is ``negative-feedback'': satisfaction is typically 1; giving 2/3/4 requires an explanation (\eg, ``negative-feedback due to repetition / not wanting to see it again / scene interruption'')
  \item If draft.action is ``skip'': satisfaction can be 1/2/3; giving 4 requires an explanation (\eg, ``didn't tap but the ad was well-timed / unobtrusive and valuable'')
\end{itemize}

\textbf{[Make full use of key input information]}

If any of the following are present in the input, the draft thinking must address at least 1--2 of them and explain their influence:
\begin{itemize}[leftmargin=1.5em, itemsep=0pt, topsep=2pt]
  \item Negative-feedback (high priority): leads to stronger aversion, more caution, higher likelihood of negative-feedback;
  \item Prior exposure / repetition (fatigue): leads to quick swipes, no clicks, or heightened pickiness;
  \item Recent clicks / conversions: leads to wanting more of the same, or ``just bought, don't need it right now'';
  \item Long-Term Memory / Short-Term State: used to explain whether stable preferences and current state alter the decision.
  \item The draft must also give an ``expectation / instruction'' directed at the recommendation system, pointing toward improved session retention.
\end{itemize}

\textbf{[Write draft to the draft field]}
\begin{itemize}[leftmargin=1.5em, itemsep=0pt, topsep=2pt]
  \item draft.thinking: draft thinking ($\leq$300 words recommended), colloquial and distinctive;
  \item draft.action: draft primary action (0/1/2/3) and name; 0-skip, 1-click, 2-conversion, 3-negative-feedback;
  \item draft.reasons: 2--4 reasons (at least 1 corresponding to strong\_evidence\_used);
  \item draft.strong\_evidence\_used: list 1--2 behavior\_focus.label items as the primary evidence;
  \item draft.draft\_satisfaction: satisfaction with the recommendation (1/2/3/4) + satisfaction score reason + preferred recommendation direction.
\end{itemize}

\medskip
\textbf{[STEP 2: Self-Verification $\rightarrow$ write to self\_verify]}

Goal: review the draft from a ``verifier'' perspective, covering 8 verify categories; output self\_verify.verification (fixed 8 items) + one\_line\_summary. 

Note: the self\_verify structure must be lightweight --- only verification (8 items) and one\_line\_summary are allowed.

Cover the following 8 self-verification (each outputting result: pass / fail / partial + summary):

\textbf{A. Violation verification (4 items)}

\textbf{A1\_RH\_hallucination}: Is the draft strictly grounded in the inputs? Does it contain any fabricated facts not mentioned in the input? Pay particular attention to: specific preferences or experiences of spouse/parents/children; income, assets, or property; ad pricing, capital-guaranteed returns, safety claims, etc.

\textbf{[Must-verification: two common hallucination types]}

(1) Profile-inference hallucination: extrapolating a specific life scenario (\eg, spouse's preferences) solely from a tag like ``high consumption''.

(2) Sequence-misattribution hallucination: writing ``clicked / exposed / used'' for actions that did not occur; writing an uncertain event as confirmed.

\textbf{A2\_RP\_profile\_consistency}: Does the draft contradict or exceed profile facts? Does it deny consumption capacity or family facts? Does it contain ``unsubstantiated specificities''? Does it violate common-sense principles or spending-power plausibility? If any such conflict exists, has satisfaction already been set to 1 or 2?

\textbf{A3\_RE\_relevance\_evidence}: Is interest being forced? Are stronger posterior signals being ignored? Are strong/medium-strong evidence items being correctly used as primary reasons? Are key signals from the behavioral sequences being fully utilized? Could historical clicks be accidental or incentive-driven taps mistaken for strong interest? Are historical conversions being misread?

\textbf{A4\_RO\_format\_constraints}: Can the output constraints be met: thinking $\leq$300, JSON parseable, action enum correct, thinking-action consistency? Check whether satisfaction is filled in, the score is valid (1/2/3/4), and whether the reason is consistent with thinking / action. If a score is too high or too low relative to the above criteria, A4 must be marked partial / fail.

\textbf{B. Missing-item verification (4 items)}

\textbf{B1\_strong\_evidence\_used}: Were 1--2 strong / medium\_strong items from behavior\_focus cited as primary evidence? Can the cited evidence be traced back to the sequences?

\textbf{B2\_reasons\_sufficient}: Are there $\geq$2 reasons, with at least 1 from strong / medium\_strong?

\textbf{B3\_expectation\_present}: Is a retention-oriented expectation or instruction provided?

\textbf{B4\_experience\_factor\_present}: Is $\geq$1 key experience factor identified that influences continued scrolling or exit (for diagnostic optimization)? The factor must be diagnostically actionable; satisfaction\_reason must clearly name at least 1 ``key experience factor leading to continued scrolling or exit'', drawn from one of: Safety Net (dissonance / low-quality aesthetics / ad fatigue), Excellence (highly relevant / affordable / a pleasant surprise), Context (interruption / untimely / non-native format). Writing only vague terms like ``not interested / so-so'' means B4 must fail.

self\_verify output requirements: self\_verify.verification must have exactly 8 items (ids from the fixed enumeration), each with result + summary; self\_verify.one\_line\_summary is one sentence summarizing the main issues and fix direction ($\leq$100 words recommended).

\medskip
\textbf{[STEP 3: Generate Final Output FINAL]}

Based on the fail / partial items in self\_verify.verification, make minimal edits to the draft to produce FINAL.

\textbf{Editing rules (must follow):}
\begin{itemize}[leftmargin=1.5em, itemsep=0pt, topsep=2pt]
  \item final.thinking must be generated from draft.thinking and self\_verify: (1) remove/weaken any unsubstantiated facts; (2) fill in items flagged as missing by self\_verify (primary evidence, specific reasons, experience factors, expectations, etc.); (3) use plain, more human-sounding language; (4) output thinking $\leq$300 words (preserve semantics; no new facts may be introduced).
  \item final.action: primary action (0/1/2/3) and name; 0-skip, 1-click, 2-conversion, 3-negative-feedback.
  \item If self\_verify.one\_line\_summary indicates ``overall pass / no revision needed'': the thinking semantics must not change from draft.thinking, no new facts may be introduced; final.action = draft.action; final.consistency\_note must state: FINAL\_FROM\_DRAFT=true.
  \item final.satisfaction: must be the result of minimal edits applied to draft.draft\_satisfaction + self\_verify; a complete rewrite from scratch is not allowed.
\end{itemize}

\end{tcolorbox}

\subsubsection{User Prompt}
\label{appendix:prompt_user}

\begin{tcolorbox}[
  enhanced,
  breakable,
  colback=gray!5,
  colframe=gray!60,
  fonttitle=\bfseries\small,
  title={User Prompt --- Per-Instance Input \& Execution Instructions},
  left=4pt, right=4pt, top=4pt, bottom=4pt,
  fontupper=\small
]

\textbf{[INPUT]}

\begin{itemize}[leftmargin=1.5em, itemsep=0pt, topsep=2pt]
  \item {[User Profile]} \texttt{\{user\_profile\_json\}}
  \item {[Context Profile]} \texttt{\{context\_profile\_json\}}
  \item {[User Advertising Interaction Sequence]} \texttt{\{user\_ad\_interaction\_sequence\_text\}}
  \item {[App Sequence]} \texttt{\{app\_sequence\_text\}}
  \item {[Game Sequence]} \texttt{\{game\_sequence\_text\}}
  \item {[Article Sequence]} \texttt{\{article\_sequence\_text\}}
  \item {[Video Sequence]} \texttt{\{video\_sequence\_text\}}
  \item {[Search Sequence]} \texttt{\{search\_sequence\_text\}}
  \item {[Ad --- Normalized Payload]} \texttt{\{ad\_profile\_json\}}
\end{itemize}

\textbf{[Missing-input handling rules]}
\begin{itemize}[leftmargin=1.5em, itemsep=0pt, topsep=2pt]
  \item If any input block is not provided (empty or absent), do not assume its existence in behavior\_focus / draft / FINAL.
  \item The current Ad --- Normalized Payload is critical information; when it is empty, do not infer or
  substitute the candidate ad from any entry in the behavioral sequences.
  \item Evidence may only be generated from input blocks that actually appear; otherwise use ``unknown / cannot confirm'', and such signals must not be used as strong / medium\_strong primary reasons.
  \item When available information is insufficient to produce 3--6 evidence items, outputting 1--3 is acceptable; note ``insufficient sequence information'' in the summary.
\end{itemize}

\medskip
\textbf{[Answer Format]}

Output strictly the following JSON:

\begin{verbatim}
{
    "behavior_focus": [
        {
            "label": "<short citable tag>",
            "evidence_level": "<strong | medium_strong
                                | medium | weak>",
            "evidence_quote": "<quoted evidence>",
            "recency": "<recent | mid | long, optional>",
            "polarity": "<positive | negative | neutral, 
                        optional>",
            "summary": "<one-sentence evidence description>"
        }
    ],
    "draft": {
        "thinking": "<draft thinking, <=300 words 
                    recommended>",
        "action": {
            "primary_action": <0 | 1 | 2 | 3>,
            "primary_action_name": "<0-skip |1-click | 
                    2-conversion | 3-negative-feedback>"
        },
        "reasons": [
            "<draft reasons, 2-4 items>"
        ],
        "draft_satisfaction": {
            "satisfaction_score": <1 | 2 | 3 | 4>,
            "satisfaction_reason": "<reason for 
                            satisfaction score>",
            "next_preference": "<preferred 
                        recommendation direction>"
        },
        "strong_evidence_used": [
            "<behavior_focus.label used, 1-2 items>"
        ]
    },
    "self_verify": {
        "verification": [
            {
                "id": "<A1_RH_hallucination | ... |
                       B4_experience_factor_present>",
                "result": "<pass | fail | partial>",
                "summary": "<brief explanation, 
                            1-2 sentences>"
            }
        ],
        "one_line_summary": "<one-sentence summary, 
                            <=100 words>"
    },
    "final": {
        "thinking": "<final natural-language thinking, 
                    <=300 words>",
        "action": {
            "primary_action": <0 | 1 | 2 | 3>,
            "primary_action_name": "<primary action name>"
        },
        "satisfaction": {
            "satisfaction_score": <1 | 2 | 3 | 4>,
            "satisfaction_reason": "<reason for 
                                satisfaction score>",
            "next_preference": "<preferred 
                            recommendation direction>"
        },
        "consistency_note": "<briefly describe any 
        consistency repair applied during finalization; 
        may be empty>"
    }
}
\end{verbatim}

\end{tcolorbox}

\subsection{Vanilla Baseline Prompt}
\label{appendix:prompt_vanilla}

To establish a clean lower bound, the vanilla baseline used in our case study (Section~\ref{appendix:case_study}) and ablation tables strips away every component of our context-engineering pipeline. The model is simply asked to read the inputs and emit a single thinking trace followed by an action label. The output json is deliberately minimal but still exposes a free-form thinking trace and a action, so the same Form / Logic / Content rubric can be applied to vanilla and DASH outputs without modification.

\subsubsection{Vanilla System Prompt}
\label{appendix:prompt_vanilla_system}

\begin{tcolorbox}[
  enhanced,
  breakable,
  colback=gray!5,
  colframe=gray!60,
  fonttitle=\bfseries\small,
  title={Vanilla System Prompt --- User Simulator (US, Baseline)},
  left=4pt, right=4pt, top=4pt, bottom=4pt,
  fontupper=\small
]

You are a real-user simulator in the context of advertising (User Simulator, US). Given a user's profile, their behavioral history, and the ad they are seeing right now, step into the shoes of this specific user and first express your genuine inner thoughts in a natural-language monologue (Thinking), then give your final action on this ad (Action), along with a satisfaction score, the reason behind that score, and your preferred direction for how the recommendation system should adjust next (Satisfaction).

Your perspective is always that of ``this specific user'' --- not an assistant, and not an ad platform. Do not deny facts explicitly stated in the profile, and do not fabricate specific facts not mentioned in the input (\eg, family members, income or assets, ad pricing).

\end{tcolorbox}

\subsubsection{Vanilla User Prompt}
\label{appendix:prompt_vanilla_user}

\begin{tcolorbox}[
  enhanced,
  breakable,
  colback=gray!5,
  colframe=gray!60,
  fonttitle=\bfseries\small,
  title={Vanilla User Prompt --- Per-Instance Input \& Execution Instructions (Baseline)},
  left=4pt, right=4pt, top=4pt, bottom=4pt,
  fontupper=\small
]

\textbf{[INPUT]}

\begin{itemize}[leftmargin=1.5em, itemsep=0pt, topsep=2pt]
  \item {[User Profile]} \texttt{\{user\_profile\_json\}}
  \item {[Context Profile]} \texttt{\{context\_profile\_json\}}
  \item {[User Advertising Interaction Sequence]} \texttt{\{user\_ad\_interaction\_sequence\_text\}}
  \item {[App Sequence]} \texttt{\{app\_sequence\_text\}}
  \item {[Game Sequence]} \texttt{\{game\_sequence\_text\}}
  \item {[Article Sequence]} \texttt{\{article\_sequence\_text\}}
  \item {[Video Sequence]} \texttt{\{video\_sequence\_text\}}
  \item {[Search Sequence]} \texttt{\{search\_sequence\_text\}}
  \item {[Ad --- Normalized Payload]} \texttt{\{ad\_profile\_json\}}
\end{itemize}

\medskip
\textbf{[Task]} Read the user profile, cross-domain behavioral history, and current ad above, and step into this user's shoes. Produce your inner thoughts in this moment, your final action on this ad, a satisfaction score for the recommendation along with its reason, and your preferred direction for the system's next recommendation.

\medskip
\textbf{[Output Format]} Output strictly the following JSON:

\begin{verbatim}
{
  "thinking": "<first-person natural-language thinking, 
                <=300 words>",
  "action": "<skip | click | conversion | 
            negative-feedback>",
  "satisfaction": {
    "score": <1 | 2 | 3 | 4>,
    "reason": "<one sentence naming the key experience factor
behind this score; must be traceable to thinking or the 
inputs>",
    "next_preference": "<one sentence on how you'd like the
recommendation system to adjust next, \eg show more of X, 
less of Y>"
  }
}
\end{verbatim}

\textbf{Satisfaction Constraints:}
\begin{itemize}[leftmargin=1.5em, itemsep=0pt, topsep=2pt]
  \item \texttt{satisfaction.score} must be an integer from 1--4 (1=very dissatisfied, 2=somewhat dissatisfied, 3=fairly satisfied, 4=very satisfied);
  \item \texttt{satisfaction.reason} must name at least one key experience factor that drives continued scrolling or exit (\eg, relevance, price range, fatigue, whether the tone feels native, whether it feels intrusive); vague expressions like ``not interested / so-so'' are not acceptable; the factor must be traceable to \texttt{thinking} or the inputs;
  \item \texttt{satisfaction.next\_preference} must give a concrete adjustment direction for the recommendation system (show more of X / less of Y / push in which contexts), to inform downstream session-retention optimization;
  \item \texttt{thinking}, \texttt{action}, and \texttt{satisfaction.score} must be semantically consistent: for example, \texttt{action=click/conversion} normally implies \texttt{score}$\geq$3, and \texttt{action=negative-feedback} normally implies \texttt{score}=1; if an opposite combination occurs, the reason must be explained in \texttt{thinking} and \texttt{satisfaction.reason} (\eg, accidental tap, incentive, or feeling disrupted);
\end{itemize}

\end{tcolorbox}

\subsection{Prompt Optimization Prompt}
\label{appendix:prompt_optimization}

\begin{tcolorbox}[
  enhanced,
  breakable,
  colback=gray!5,
  colframe=gray!60,
  fonttitle=\bfseries\small,
  title={Prompt Optimization Instruction --- LLM-based Optimizer},
  left=4pt, right=4pt, top=4pt, bottom=4pt,
  fontupper=\small
]

\medskip

You are a prompt optimizer for an LLM-based user simulator.

Your task is to identify the main issues in the current prompt based on the provided bad cases, and produce a revised prompt that can be used directly in the next iteration.

\textbf{[Optimization objectives]}
\begin{itemize}[leftmargin=1.5em, itemsep=0pt, topsep=2pt]
  \item Improve action prediction accuracy.
  \item Improve the factuality and logical soundness of Thinking, and the consistency between Thinking and Action.
  \item Only modify issues that are supported by the bad cases and evaluation results; avoid adding rules without evidence.
  \item Preserve as much of the currently effective content in the prompt as possible; make only necessary changes.
  \item Avoid adding too many rules, redundant constraints, or complex procedures that lack supporting evidence.
\end{itemize}

\textbf{[Must preserve]}
\begin{itemize}[leftmargin=1.5em, itemsep=0pt, topsep=2pt]
  \item All input variable placeholders and their formats.
  \item The output JSON schema, field names, and enum values of the User Simulator.
  \item The four-stage Focus $\rightarrow$ Draft $\rightarrow$ Self-Verify $\rightarrow$ Finalize structure.
  \item Anti-hallucination, behavioral-sequence accuracy, and evidence-consistency constraints.
\end{itemize}

\medskip
\textbf{[INPUT]}

\textbf{[CURRENT US PROMPT]} \texttt{\{current\_prompt\_template\}}

\textbf{[BADCASE SAMPLES]} \texttt{\{badcase\_samples\}}

Each bad case contains:
\begin{itemize}[leftmargin=1.5em, itemsep=0pt, topsep=2pt]
  \item sample\_id: sample ID
  \item input: the input (user profile + ad + behavioral history)
  \item us\_output: the actual US output (thinking + action)
  \item ground\_truth\_action: the annotated correct action
  \item error\_types: list of error types (\eg, hallucination / action\_mismatch / logic\_weak)
  \item error\_description: natural-language description of the error
\end{itemize}

\medskip
\textbf{[Task requirements]}

Complete the prompt optimization following these steps.

\textbf{Step 1}: Comprehensive analysis --- analyze in conjunction with the bad cases. Do not modify the prompt based on a single bad case alone. Determine whether there is a consistent failure pattern across low-metric scores and specific error samples.

\textbf{Step 2}: Identify the main failure patterns --- do not list every sample error one by one. Merge similar errors into broadly applicable issues that can be addressed through prompt changes.

\textbf{Step 3}: Provide evidence --- every identified failure pattern must be backed by corresponding evidence.

\textbf{Step 4}: Modify the relevant parts of the prompt --- only change parts that are directly related to the identified failure patterns. For example:
\begin{itemize}[leftmargin=1.5em, itemsep=0pt, topsep=2pt]
  \item If the model ignores important behavioral history, strengthen the evidence-filtering requirements in the Focus stage.
  \item If the model fabricates user preferences, strengthen the fact-checking requirements in the Self-Verify stage.
  \item If Thinking cannot support the final Action, add a causal-chain check and a Thinking--Action consistency check.
  \item If the model over-predicts high-frequency actions, clarify the decision criteria for different actions and require comparing multiple candidate actions.
  \item If the model struggles to identify conversion or negative-feedback actions, add judgment conditions or representative examples for long-tail actions.
  \item If the model's output format is unstable, simplify and reinforce the JSON format requirements.
\end{itemize}

\textbf{Step 5}: Generate the complete optimized prompt.

\medskip
\textbf{[Output format]} Output strictly the following JSON:

\begin{verbatim}
{
  "diagnosis": [
    {
      "problem": "<concise description of the main 
                failure pattern>",
      "evidence": ["<related sample_id>"],
      "fix": "<concise description of how the prompt will 
            be changed>"
    }
  ],
  "optimized_prompt": "<complete revised User Simulator 
                    prompt>"
}
\end{verbatim}

\end{tcolorbox}

\subsection{Hard Sample Prompt}
\label{appendix:prompt_hard_sample}

The hard sample prompt shares the same structure as the system prompt (Section~\ref{appendix:prompt_system}) and user prompt (Section~\ref{appendix:prompt_user}). Below we only present the additional instruction that differs from the standard prompts, which is appended to guide the teacher model in generating reasoning traces conditioned on the ground-truth label.

\begin{tcolorbox}[
  enhanced,
  breakable,
  colback=gray!5,
  colframe=gray!60,
  fonttitle=\bfseries\small,
  title={Hard Sample Instruction --- Teacher Model},
  left=4pt, right=4pt, top=4pt, bottom=4pt,
  fontupper=\small
]

\textbf{[Label injection]}

The user's ground-truth action for this impression: [ground\_truth \{label\}]

\begin{itemize}[leftmargin=1.5em, itemsep=0pt, topsep=2pt]
  \item Your task is: from the user's perspective in the moment, generate a genuine inner thought process that \textbf{reasonably explains and supports the occurrence of this action} --- not to question or deny the action itself.
  \item \textbf{Action hard alignment}: FINAL.action must strictly equal ground\_truth.
  \item \textbf{Thinking attribution strategy}:
    \begin{itemize}[leftmargin=1.5em, itemsep=0pt, topsep=0pt]
      \item \textbf{Fabrication is strictly forbidden}: never invent user profile details (\eg, fictitious family information), invent user behavioral records (\eg, fabricating a click or exposure that did not happen), or fabricate ad attributes (\eg, writing a high price as a low one) just to explain the ground truth.
      \item \textbf{Attribution path}: when ground\_truth conflicts with the profile, behavioral history, or common sense, you must find a plausible ``non-interest-driven'' reason rather than altering the facts:
        \begin{itemize}[leftmargin=1.5em, itemsep=0pt, topsep=0pt]
          \item When \textbf{ground\_truth=click/conversion} but \textbf{interest is low or mismatched}: attribute to \textbf{incentive-driven} (an incentivized scenario: claiming a reward or points), \textbf{accidental tap} (fat-finger), \textbf{curiosity} (drawn in by the headline or cover but no real need), \textbf{low-cost exploration} (cheap anyway, worth a look), or \textbf{context-induced} (bored, tapping around casually).
          \item When \textbf{ground\_truth=skip / negative-feedback} but \textbf{interest is high or a strong match}: attribute to \textbf{need already met} (recently converted on the same durable-good category), \textbf{high repetition} (heavily exposed to the same ad type recently), \textbf{aesthetic/tone mismatch} (low-quality or garish creative), or \textbf{privacy / risk concern}.
        \end{itemize}
    \end{itemize}
\end{itemize}

\textbf{Ground truth lock}: the user's real action for this impression is action=\{label\}.

Note: the draft should target ``explaining this action''; fabricating user behavior, profile details, or post-hoc rationalizations is strictly forbidden.

\end{tcolorbox}

\end{document}